\documentstyle[12pt,epsfig,aaspp4,figwithcaption]{article} 
\newcommand{\etal}{et~al.~}
\newcommand{\Msun}{\ifmmode{M_\odot}\else$M_\odot$~\fi}
\newcommand{\kms}{\hbox{km\thinspace s$^{-1}$}}
\newcommand{\kmsMpc}{\hbox{km\thinspace s$^{-1}$\thinspace
Mpc$^{-1}$}} \newcommand{\TSPH}{{\sc TreeSPH~}}
 \newcommand{\disk}{{\rm disk}}
 
 \newcommand{\jt}{{\tilde\jmath}}
\newcommand{\be}{\begin{eqnarray}}
\newcommand{\ee}{\end{eqnarray}}
\newcommand{\bi}{\bibitem}

\newcommand{\rar}{\rightarrow}

\newcommand{\num}{\nu_\mu}
\newcommand{\nut}{\nu_\tau}

\begin{document}
\title{Formation of Disk Galaxies:\\ Warm Dark Matter and the Angular Momentum
problem} \author{Jesper Sommer-Larsen and Alexandre Dolgov}
\affil{ Theoretical Astrophysics Center,
Juliane Maries Vej 30, DK-2100 Copenhagen {\O}, Denmark}

\begin{abstract}
We have performed \TSPH simulations of disk galaxy formation in various warm 
dark matter (WDM) cosmologies. Our results indicate that for a range of
WDM free-streaming masses, the disk galaxy
formation angular momentum problem can be considerably alleviated (and we
speculate: perhaps even completely resolved) by
going to the WDM structure formation scenario, without having to invoke
stellar feedback processes. They also strongly suggest
that part of the angular momentum problem is due to numerical effects, most
likely related to the shock capturing, artificial viscosity used in SPH.
Furthermore we find that we can match the observed $I$-band Tully-Fisher (TF)
relation, provided that the mass-to-light ratio of disk galaxies is
$(M/L_I) \simeq$ 0.6-0.7. We argue that this is a fairly reasonable value
in comparison with various dynamical and spectrophotometric estimates,
including two given in this paper. 
Finally, we discuss possible physical candidates for WDM particles extensively.
We find that
the most promising are neutrinos with weaker or stronger interactions than
normal, majorons (light pseudogoldstone bosons) or mirror or shadow world
neutrinos.
\end{abstract}
 
\section{Introduction}
\label{s:intro}

The formation of galactic disks is one of the most important unsolved
problems in astrophysics today. In the currently favored hierarchical
clustering framework, disks form in the potential wells of dark matter
halos as the baryonic material cools and collapses dissipatively.  It
has been shown (Fall \& Efstathiou \cite{FE80}) that
disks formed in this way can be expected to possess the observed
amount of angular momentum (and therefore the observed spatial extent
for a given mass and profile shape), but only under the condition that the
infalling gas retain most of its original angular momentum.

Numerical simulations of this collapse scenario in the cold dark matter
cosmological context 
(e.g., 
Navarro \& Benz \cite{NB91},
Navarro \& White \cite{NW94},
Navarro, Frenk, \& White \cite{NFW95}), 
however,
have so far consistently indicated that when only cooling processes are
included the infalling gas loses too much angular momentum (by over
an order of magnitude) and the resulting disks are accordingly much
smaller than required by the observations.
This discrepancy is known as the {\em angular momentum problem} of
disk galaxy formation.
It arises from the combination of the following two facts:
a) In the CDM scenario the magnitude of linear density fluctuations  
$\sigma(M) = <(\delta M/M)^2>^{1/2}$ increases steadily with decreasing
mass scale $M$ leading to the formation of non-linear, virialized structures
at increasingly early epochs with decreasing mass i.e.~the hierarchical
``bottom-up'' scenario.  b) Gas cooling
is very efficient at early times due to gas densities being generally
higher at high redshift as well as the rate of inverse Compton cooling also 
increasing very rapidly with redshift ($\propto (1+z)^4$).  a) and b)
together lead to rapid condensation of small, dense gas clouds,
which subsequently lose energy and (orbital) angular momentum by dynamical
friction against the surrounding dark matter halo before they
eventually merge to form the central disk.
A mechanism is therefore needed that prevents, or at least delays,
the collapse of small protogalactic gas clouds and allows the gas to
preserve a larger fraction of its angular momentum as it settles into
the disk.
(Such a mechanism is also helpful in solving the {\em overcooling
problem}, namely the observation (White \& Rees \cite{WR78})
that cooling is expected to be so efficient at early times that most
of the gas should have been converted to stars well before the
assembly of present-day galactic disks).
Weil, Eke, \& Efstathiou (\cite{WEE98}) 
have shown that if the early
cooling is suppressed (by whatever means), numerical simulations can
indeed yield more realistically sized disks - see also Eke, Efstathiou \&
Wright (\cite{EEW99}).
The physical mechanism by which cooling is suppressed or
counteracted, however, was not specified.

Sommer-Larsen \etal (\cite{S.99}, hereafter SLGV99) discussed the effects of 
various
stellar reheating mechanisms in more detail using numerical
\TSPH simulations of disk galaxy formation in a CDM cosmological
context. They found that more or less uniform reheating of the Universe 
resulting
from a putative early epoch of population III star formation ($z \ga 6)$
does not lead to a solution to the angular momentum problem, but that localized
star-bursts in protogalactic gas clouds might: {\it If} the star-bursts can
blow the remaining, bulk part of the gas out of the small
and dense dark matter halos of the clouds, then the test simulations of
SLGV99 show that the gas later gradually
settles forming an extended, high angular momentum disk galaxy in the
central parts of a large, common dark matter halo. 
The physics of global gas blow-out processes were considered in early 
calculations by
Dekel \& Silk (\cite{DS86}) and Yoshii \& Arimoto (\cite{YA87}) indicating
that star-bursts might well blow out most of the gas in small galaxies with
characteristic circular speed (where the rotation curve is approximately
constant) $V_c \la$ 100 km/s. Unfortunately, more recent, detailed simulations
by Mac Low \& Ferrara
(\cite{MF99}) suggest that this global blow-out scenario may not work so
well, even in small (disk) galaxies: The star-bursts typically lead to bipolar
outflows of very hot gas perpendicular to the disk of the small galaxy,
only expelling a minor fraction of the disk gas.  

Moreover, various possible shortcomings of the CDM cosmological scenario 
in relation to structure formation on galactic scales have recently been 
discussed
in the literature: 1) CDM possibly leads to the formation of too many
small galaxies relative to what is observed, i.e. the {\em missing satellites
problem} (e.g., Klypin \etal \cite{K.99}).  2) Even if galactic winds due
to star-bursts can significantly reduce the number of visible dwarf
galaxies formed, sufficiently many of the small and tightly bound dark matter 
systems left behind can still survive to the present day in the dark matter
halos of larger galaxies like the Milky Way to possibly destroy the large,
central disks via gravitational heating, as discussed by Moore \etal 
\cite{M.99a}.  3) The dark matter halos produced in CDM cosmological
simulations tend to have central cusps with $\rho_{DM}(r) \propto r^{-N},
N \sim 1-2$ (Dubinski \& Carlberg \cite{DC91}, Navarro \etal \cite{NFW96},
Fukushige \& Makino \cite{FM97}, Moore \etal \cite{M.98}, Kravtsov \etal 
\cite{K.98}, Gelato \& 
Sommer-Larsen \cite{GSL99}). This is in disagreement with the flat, central
dark matter density profiles (cores) inferred from observations of the 
kinematics of dwarf and low surface brightness galaxies (e.g., Burkert 
\cite{B95}, de Blok \& McGaugh \cite{dBMG97},
Kravtsov \etal \cite{K.98}, Moore \etal \cite{M.99b}, but see also 
van den Bosch \etal \cite{B.99}).

The first two problems may possibly be overcome by invoking warm dark
matter (WDM) instead of CDM: WDM is similar to CDM on mass scales larger
than the free-streaming mass $M_{f,WDM} \sim$ 10$^{10}$-10$^{12}$ \Msun,
but density fluctuations $\sigma_{WDM}(M)$ are suppressed on mass scales
$M < M_{f,WDM}$ relative to CDM (but note that they are still non-zero as
$\sigma_{WDM}(M<M_{f,WDM}) \simeq \sigma_{WDM}(M_{f,WDM})$ - see 
section~\ref{s:WDM}).
As a consequence, fewer low mass galaxies (or ``satellites'') are formed
cf., e.g., Moore \etal (\cite{M.99a}) and section~\ref{s:WDM} of this paper. 
The 
central cusps problem may be more generic cf. Huss \etal (\cite{H.99}) and
Moore \etal (\cite{M.99b}),
but the WDM scenario deserves further attention also on this point.
In particular the fact the momentum part of the phase-space distribution
function is not approximately singular at redshifts where all pertubations
are still in the linear regime, but has a finite width (or equivalently:
velocity dispersion) due to free-streaming may be very helpful in solving
the central cusps problem, as recently discussed by Hogan (\cite{H99}) -
see also Hogan \& Dalcanton (\cite{HD00}), Dalcanton \& Hogan (\cite{DH00})
and Madsen (\cite{M00}).

In this paper we show that the disk galaxy formation angular momentum
problem can be considerably alleviated by invoking the WDM rather
than the CDM structure formation scenario and this without having to
appeal to stellar feedback processes.
The main reason for this is that the inflow of
gas onto the forming disk is more smooth and coherent for WDM than for
CDM, resulting in considerably larger disk angular momenta, as discussed
by SLGV99.
    
In section~\ref{s:WDM} WDM and its relation to galaxy formation is briefly
discussed. Section~\ref{s:methods} gives a short presentation of the 
numerical code 
and the initial conditions. The simulations themselves are
described in section~\ref{s:simulations}, and the results are analyzed in
section~\ref{s:results}. In section~\ref{s:candidates} we discuss 
possible
physical candidates for WDM particles extensively and in 
section~\ref{s:conclusions}
we present a final discussion and summarize our conclusions.

\section{Warm Dark Matter}
\label{s:WDM}
The terms ``warm'' and ``cold'' dark matter were probably first used in
Primack \& Blumenthal (\cite{PB83}) and the implications for structure
formation theory first worked out by Blumenthal \etal (\cite{B.82}). 
In this section we give a brief introduction to ``conventional'' warm dark 
matter following the approach of Bardeen \etal (\cite{B.86}) and using 
expressions from that work - see also, e.g., Primack (\cite{P81}),
Pagels \& Primack (\cite{PP82}) and Pierpaoli \etal (\cite{P.98}).
We defer a much more detailed and general discussion of the relevant
particle physics to section~\ref{s:candidates}.

At low redshift warm dark matter particles behave in many respects in the
same way as cold dark matter (see below), but from a particle physics 
point of view they are quite different, however: CDM particles (apart
from axions) are thought to be very massive, normally with $m \ga $ 1 GeV, 
and to be non-relativistic when they decouple 
from the rest of the particles in the Universe at high redshift. WDM
is particles with typical masses $m \sim$ 1 keV, which in conventional theory
decouple at high
redshifts $z_{dec} \ga 10^{13}$ (or $T_{dec} \ga$ 1 GeV) being still
ultra-relativistic. At $z_{nr} \sim 10^6$-$10^7$ the WDM particles become
non-relativistic: $3 k T_{WDM} \simeq m_{WDM} c^2$. The Universe is still
radiation dominated at this stage, as $z_{nr} >> z_{eq} \sim 10^{4}$, where
$z_{eq}$ is the redshift of matter-radiation equivalence. The horizon mass
of WDM at $z_{nr}$ defines a characteristic mass scale called the 
free-streaming mass $M_{f,WDM} \sim M_{H,WDM}(z_{nr})$. 
Pertubations on mass scales
$M \la M_{f,WDM}$ are damped relative to CDM due to relativistic 
free-streaming of the WDM particles at $z > z_{nr}$. On larger mass scales 
$M \ga M_{f,WDM}$ WDM behaves like CDM.

Following Bardeen \etal (\cite{B.86})
the power spectrum of WDM (adiabatic fluctuations) 
can be expressed as
\begin{equation}
P_{WDM}(k) = T_{WDM}^2(k)~P_{CDM}(k) ~~,
\end{equation}
where the CDM to WDM ``transfer function'' can be approximated well by 
\begin{equation}
T_{WDM}(k) = \exp\left[-\frac{k R_{f,WDM}}{2} -\frac{(k R_{f,WDM})^2}{2} 
\right]  
\end{equation}
and $P_{CDM}(k)$ is the CDM power spectrum for which we use the standard
SCDM form. The function $T_{WDM}^2$ is shown in Figure 1. 
The comoving, free-streaming scale $R_{f,WDM}$ is given by
\begin{equation}
R_{f,WDM} = 0.2~\left(\frac{g_{WDM,dec}}{100}\right)^{-4/3}~
(\Omega_{WDM}h^2)^{-1} ~\mbox{Mpc} ~~, 
\end{equation}
$g_{WDM,dec}$ being the effective number of particle degrees of freedom
when the WDM particles decouple. $\Omega_{WDM}h^2$ is related to the
mass of the WDM particles by
\begin{equation}
\Omega_{WDM}h^2 = 1.0~\left(\frac{g_{WDM,dec}}{100}\right)^{-1}~
\left(\frac{m_{WDM}}{\mbox{keV}}\right)
\end{equation}
For justification of equations (3) and (4), see section~\ref{s:candidates}.

A characteristic free-streaming wave number $k_{f,WDM}$ can be defined as the 
$k$ for which $T_{WDM}^2$=0.5. From Figure 1 it then follows that $k_{f,WDM} 
\times R_{f,WDM} \simeq$ 0.46, so we define $k_{f,WDM} \equiv R_{f,WDM}/0.46$.
Using this we can then define a characteristic free-streaming mass in terms of
$\lambda_{f,WDM} \equiv 2 \pi/k_{f,WDM}$ by
\begin{equation}
M_{f,WDM} \equiv \frac{4 \pi}{3} ~\rho_{crit} ~\Omega_{WDM} ~\left(\frac
{\lambda_{f,WDM}}{2}\right)^3 = 
3.7\cdot 10^{11} ~h^{-1} ~\Omega_{WDM} ~\left(\frac{R_{f,WDM}}{0.1 h^{-1} \mbox{Mpc}}\right)^3 ~\mbox{M}_{\odot}
\end{equation}
Using equations (3)-(5) we can finally express the WDM particle mass in terms 
of this free-streaming mass  
\begin{equation}
m_{WDM} = 2.4 ~h^{5/4} ~\Omega_{WDM}^{1/2} ~\left(\frac{M_{f,WDM}}{10^{11} 
h^{-1} \Msun}\right)^{-1/4} ~\mbox{keV} ~~.
\end{equation}

The average relative mass fluctuations on mass scale $M$, $\sigma(M)$, 
can be calculated from the power spectrum $P(k)$ in linear theory as
\begin{equation}
\sigma^2(M) = ~<\left(\frac{\delta M}{M}\right)^2> ~= ~\frac{1}{(2 \pi)^3}
\int d^3 k ~P(k) ~W(k x) ~~,
\end{equation}
where the weight function
\begin{equation}
W(y) = \frac{9}{y^6} [\sin y - y \cos y]^2 ~, ~~y = k x
\end{equation}
is the square of the Fourier transform of a spherical top-hat filter of radius
$x$ (following Peebles \cite{P80}). 
To illustrate the difference between CDM and WDM we show in Figure 2 
$\sigma(M;z=0)$ for CDM as well as WDM with $R_{f,WDM}$ = 0.037,
0.075 and 0.15 $h^{-1}$Mpc corresponding to free-streaming masses of
$1.9\times 10^{10}, ~1.5\times 10^{11}$ and $1.2\times 10^{12} 
~~h^{-1} \mbox{M}_{\odot}$ The curves have been normalized such that 
$\sigma_8$ = 0.5, where $\sigma_8$ is the present day value of $\sigma(M)$ 
defined above for $M$ equal to the average mass within spheres of comoving
radius 8$h^{-1}$ Mpc.     

The fact that the $\sigma(M;z=0)$ curves flatten at low masses for WDM does
{\em not} mean that no low mass galaxies are formed (see also Schaeffer \&
Silk \cite{SS88}). To illustrate this
we calculate the present day ($z=0$) mass spectrum using (linear) 
Press-Schechter (PS) theory: The mass spectrum $dN(M)/dM$ in the linear
PS approximation is given by (e.g. White \cite{W93})
\begin{equation}
\frac{dN(M)}{dM}(z=0) = -\sqrt{\frac{2}{\pi}} \frac{\bar{\rho}}{M}
\frac{\delta_c}{\sigma^2(M;z=0)} \frac{d\sigma(M;z=0)}{dM} \exp\left[
\frac{-~\delta_c^2}{2 \sigma^2(M;z=0)}\right] ~~,
\end{equation}
where we take $\delta_c$ = 1.69 (see White \cite{W93}).
The mass spectra obtained in this way for CDM and the 
WDM described above are shown in Figure 3. Clearly, for WDM $dN/dM$ still 
decreases with $M$ as for CDM, but, for example, for masses three orders of 
magnitude below the free-streaming mass $dN/dM$ is down by more than an
order of magnitude relative to CDM, whereas for masses of the order of or
larger than the free-streaming mass $dN/dM$ is essentially the same for
WDM and CDM. We caution, however, that the PS theory applied above is not
completely rigorous and hence that N-body WDM simulations should be undertaken
to determine WDM mass spectra more properly.

\section{The code and the initial conditions}
\label{s:methods}

\subsection{The code}
\label{s:code}

We use the gridless Lagrangian $N$-body and Smoothed Particle
Hydrodynamics code {\TSPH} described in SLGV99.
Our {\TSPH} code is modeled after that of Hernquist \& Katz (\cite{HK89}).

We include gas cooling and heating terms as in Vedel \etal (\cite{VHS94}). 
The heating corresponds to a redshift-dependent, homogeneous and isotropic UVX
background field. We assume a rather hard (spectral index $-1$) quasar like 
field
\begin{equation}
J_{\nu}(z) = J_{-21}(z) \times 10^{-21}
\left(\nu\over\nu_L\right)^{-1}
\,\hbox{erg}\,\hbox{cm}^{-2}\,\hbox{sr}^{-1}\,\hbox{Hz}^{-1}\,\hbox{s}^{-1},
\end{equation}
where $\nu_L$ is the Lyman limit frequency,
with the redshift-dependent normalization
\begin{equation}
J_{-21}(z) = { 10 \over 1 + \left[5/(1+z)\right]^4 }
\end{equation}
of Efstathiou (\cite{Ef92}).
For greater realism one could build on the detailed study of Haardt \&
Madau (\cite{HM96}), but our adopted background field is
quantitatively not too dissimilar from theirs (once allowance is made
for our simplified spectral shape), at least at redshifts~$z\la 3$,
and should be adequate for the level of detail we can represent in
the simulations.
The code furthermore incorporates inverse Compton cooling, which is also
explicitly redshift-dependent.

SLGV99 studied the effects of star-formation in terms of the
subsequent energy and momentum feedback processes to the gas as a means
of resolving the angular momentum problem of disk galaxy formation. 
In this paper
we study the effects of going from CDM to WDM, as an alternative. In order
to separate the various effects clearly we describe in this paper what
in SLGV99 was dubbed ``passive'' simulations, i.e. simulations with no
star-formation and feedback effects. 

The smoothing length of each SPH particle is
adjusted so as to keep the number of neighbors close to~50.

\subsection{The initial conditions}
\label{s:ic}

Our cosmological initial conditions are based on a standard ($\Omega_M=1$,
$\Omega_{\Lambda}=0$) CDM model with Hubble constant
$H_0 = 100h\,\kmsMpc = 50\,\kmsMpc$. On the scales of interest to
us the effective index of the power spectrum is approximately~$-2$
($P(k) \propto k^{-2}$). Following Eke, Cole, \& Frenk
(\cite{ECF96})
we normalize the spectrum to $\sigma_8(z=0)=0.5$,
where as customary $\sigma_8^2$ is the mass variance within spheres of
comoving radius $8h^{-1}$~Mpc, extrapolated from the linear regime of
perturbation growth. We note that a $\sigma_8$ of 0.5 is somewhat lower
than the value of 0.67 usually used in what is called ``standard CDM''.

We begin by performing a large-scale simulation within a sphere of
comoving radius 40~Mpc. Individual halos are then selected from the
final state and sampled at higher resolution.
The original large-scale simulation is used to provide a tidal
field acting on the resampled halos, which are evolved separately in
a second round of simulations.

Approximately $2.5\times 10^5$ particles are initially placed on a
cubic lattice 
within the large sphere. Position and velocity perturbations
are then applied according to the Zel'dovich (\cite{Ze70})
approximation.
The perturbations consist of the superposition of $N_k \sim
4\times10^4$ plane waves sampling a Gaussian random field with
variance given by the power spectrum.
Following Navarro \& White (\cite{NW94}) we use an equal number of
waves per logarithmic interval in $k$-space. The phases of these waves are
random and only wavenumbers between the fundamental and Nyquist wavenumber
of the lattice are included.
The initial redshift (which determines the amplitude of the initial
perturbations) is $z_i\simeq 18$.

The evolution of this system to $z=0$ is then computed using a tree code. Only
gravitational forces are included in this first simulation.
In the final state we identify four virialized, isolated halos with
circular velocities between 200 and 260~{\kms} in the innermost 20~Mpc of the
simulation.
We expect that a significant fraction of such halos should host disk
galaxies similar to the Milky Way since the circular velocities are
in the same range and our halos were chosen to lie ``in the field'',
away from larger concentrations of mass.
We adopt the customary working definition of the virial radius as the
radius~$r_{200}$ of a sphere enclosing a mean density of 200 times the
critical cosmic value.
Tracing the particles in these halos back to the initial conditions,
we find that they all come from regions that fit within spheres of
comoving radius $\sim 3$~Mpc.

Each of these spheres at $z\simeq18$ is then resampled with a lattice
4 times finer in each spatial dimension in our moderate resolution (MR)
simulations and 8 times finer in our high resolution (HR) simulations. 
Each sphere contains about 7000 points of this new lattice in our MR 
simulations and about 56000 points in our HR simulations.  
We assign one dark matter (DM)
particle and two SPH particles to each of these points for the MR simulations
and one DM particle and one SPH particle to each point for the HR simulations.
The SPH
particles are positioned at random within one gas gravitational softening
length (see below) of their parent DM particle.
Note that at this redshift the DM particles are spaced by about
14~kpc for the MR simulations and 7~kpc for the HR simulations, so that each 
SPH particle is rather closely associated with its parent DM particle.
Table~\ref{t:np} lists the precise number of SPH and DM particles in
each sphere.
We generally use a baryonic mass fraction $\Omega_b=0.05$, consistent with
nucleosynthesis constraints ($0.01 h^{-2} \la \Omega_b \la 0.02
h^{-2}$), but use $\Omega_b=0.10$, which is more consistent with the
observationally determined baryonic fractions in galaxy groups and clusters,
in some of the MR simulations.  This leads to masses of $1.1\times 10^9\,
\Msun$ for the
DM and $2.9\times 10^7\,\Msun$ for the SPH particles for the MR simulations
with $\Omega_b=0.05$, $1.0\times 10^9$ and $5.8\times 10^7~\Msun$ respectively
for the $\Omega_b=0.10$ MR simulations and $1.4\times 10^8$ and $7.3\times
10^6~\Msun$ for the HR simulations, which all have $\Omega_b=0.05$
The SPH particles are assigned an initial thermal energy corresponding
to a temperature $T_i\simeq 100$~K.
In order to include small-scale power that could not be sampled in the
first simulation, we add shorter-wavelength plane waves in a way that
preserves an equal number of waves per interval in~$\log k$. Given the
variety of properties of the WDM particle candidates discussed in 
section~\ref{s:candidates} and that the purpose
of this paper is to investigate qualitatively the effects on disk galaxy
formation of going from CDM
to WDM we approximate $T_{WDM}(k)$ by a step function
\begin{equation}
\tilde{T}_{WDM}(k) = \left\{
 \begin{array}{ll}
  1& ~~~~~k \le k_c ~,\\
  0& ~~~~~k > k_c~. 
 \end{array}
\right.
\end{equation}
We find that for a given $R_{f,WDM}$, $\tilde{T}_{WDM}(k)$, with 
$k_c = 0.46 R_{f,WDM}^{-1}$ (cf. section~\ref{s:WDM}), gives a reasonable 
match to $T_{WDM}(k)$
as shown in Figure 1 (dotted line). In Figure 2 we show by the thin solid line
$\tilde{\sigma}(M;z=0)$
for $k_c = 6.2 ~h~\rm{Mpc}^{-1}$ (corresponding to $R_{f,WDM} \simeq 0.075 
~h^{-1}$Mpc). As can be seen from the figure, the match to $\sigma(M;z=0)$
for $R_{f,WDM} = 0.075 ~h^{-1}$Mpc is also reasonably good.  
The WDM simulations described in this paper are hence performed using the
original, standard CDM power spectrum used in SLGV99, modified to WDM by
\begin{equation}
\tilde{T}_{WDM}(k) = \left\{
 \begin{array}{ll}
  1& ~~~~~kR_{f,WDM} \le 0.46 ~,\\
  0& ~~~~~kR_{f,WDM} > 0.46~. 
 \end{array}
\right.
\end{equation}
In this spirit we are also able to reuse the original, background CDM
cosmological simulation for the WDM simulations presented in this work, as 
the typical WDM wavenumbers $k_c$ used here are comparable to the Nyquist 
wavenumber of the CDM cosmological simulation.

We use stored intermediate results from the large cosmological
simulation to provide a time-dependent tidal field acting on the
resampled spheres.
This is achieved by treating the original particles outside the
resampled sphere as passive, interpolating their positions between
successive snapshots of the original simulation and incorporating them
into the particle tree that the code constructs on each step for the
evaluation of gravitational forces.
Gravitational interactions between particles are softened
according to the prescription of Hernquist \& Katz (\cite{HK89}), with
softening lengths of 3 kpc for the gas particles and 10 kpc for the dark 
matter 
particles in the MR simulations, 1.5 and 5 kpc, respectively, in the HR 
simulations, and 40 kpc for the ``passive'' dark matter particles that
provide the tidal field from the original large-scale simulation. The
gravitational softening lengths are kept 
constant in physical units throughout the evolution of the system.

\section{The simulations}
\label{s:simulations}

For the WDM galaxy formation simulations we identify the four resampling 
spheres S1-S4, also used by SLGV99 for their CDM galaxy formation simulations.

It seems reasonable to assume that $M_{f,WDM} \ga 10^9 - 10^{10}
\Msun$ in order for a change from
CDM to WDM to have a significant impact on the disk galaxy formation angular
momentum problem as well as the other possible CDM problems listed in 
section~\ref{s:intro}. On the other hand,
if $M_{f,WDM} \ga 10^{13} \Msun$ the bulk of galaxy and star formation will
likely take place too late relative to the observed star-formation history
of the Universe, as discussed below (see also Pierpaoli \etal \cite{P.98}).
Consequently we initially carry out three MR simulations (with 
$\Omega_b = 0.05)$
of the formation of disk galaxy S4 with $k_c$ = 12.4, 6.2 and 3.1 $~h$ 
Mpc$^{-1}$ corresponding to free-streaming masses of $1.9\times 10^{10}$,
$1.5\times 10^{11}$ and $1.2\times 10^{12} ~~h^{-1} \Msun$ by eqs. (5)
and (13). We denote warm dark matter with these characteristics WDM1, WDM2
and WDM3 in the following and the corresponding simulations runs \#1-3. 
Figure 4 shows the specific angular momentum 
$j_{disk}$ (upper panel) and cooled out gas mass $M_{disk}$ (lower panel) of 
the central,
cold and dense, forming disk (with $R \le 30$ kpc and $n_H > 0.01$ cm$^{-3}$)
as a function of time since Big Bang 
(at times when central merging is in process, i.e. when a merging
satellite is inside of $r$ = 30 kpc, $j_{disk}$ and $M_{disk}$ are not shown in
this and the following figures. Such merging episodes are clearly seen in the
figures as steep increases in $M_{disk}(t)$). The observed value
of $j$ for Milky Way sized disk galaxies is $j_{obs} \sim$ 1000-1500 kpc km/s
- see section~\ref{s:results}. 
As can be seen from Figure 4, the CDM and WDM3 simulations generally have
$j \la 200$ kpc km/s. Moreover, for the WDM3 simulations cold, dense gas, and
hence the basis for star formation, is not formed
before $t \sim$ 4.5 Gyr, corresponding to a redshift $z \sim$ 1, in 
disagreement with the observed star-formation history of the Universe - see,
e.g., Madau \etal (\cite{M.96}) and Steidel \etal (\cite{St.99}). The WDM1 and
WDM2 simulations look much more promising, both having $j \ga$ 400 kpc km/s
and for WDM2 even $j >$ 800 kpc km/s at present ($z$=0). Moreover, for the
WDM1 and WDM2 simulations, gas cools out approximately as early and at the
same rate as for the CDM simulation, but as the inflow of gas is more smooth
and coherent, considerably larger angular momenta result, as discussed by
SLGV99.

Guided by the results above we secondly carry out three MR WDM2 simulations 
(with $\Omega_b = 0.05)$
of the formation of the disk galaxies S1-S3 (runs \#4-6). As discussed
in SLGV99 one may expect that part of the angular momentum problem is related
to numerical effects, in particular angular momentum transport due to 
artificial viscosity. One would expect this effect to be reduced as the
resolution of the simulations is increased. Consequently, we thirdly carry
out four WDM2 HR simulations (with 
$\Omega_b = 0.05)$ of galaxies S1-S4 with 8 times the mass
resolution (4 times for SPH) of the MR simulations (runs \#7-10).

The effects of a 
background UVX radiation field on disk galaxy formation has been discussed
by, e.g., Vedel \etal (\cite{VHS94}) and Navarro \& Steinmetz (\cite{NS97}). 
To study the
effect of not including a UVX field in the simulations (but still inverse
Compton cooling) we fourthly carry out four MR WDM2 simulations (with 
$\Omega_b = 0.05)$ of galaxies
S1-S4 without a UVX field (runs \#11-14). Finally, to study the effect of
having a baryonic fraction of 0.10 rather than 0.05 we carry out four
MR WDM2 simulations of galaxies S1-S4 with $\Omega_b = 0.10$ and with a
background UVX radiation field (runs \#15-18). 

Parameters for all the simulations are listed in Table~\ref{t:np}.

\section{Results}
\label{s:results}

\subsection{Disk masses and specific angular momenta}

Table~\ref{t:mrv} 
presents some global properties of the final collapsed objects
in our 18 simulations. The first column is the simulation label,
as described in the previous section. Columns 2, 3, and~4 respectively
hold the virial mass $M_{200}$, the virial radius $r_{200}$, and the
circular velocity $V_{200}$ at the virial radius.
The numbers $N_{\rm gas}$ and~$N_{\rm DM}$ of gas and DM particles
inside the virial radius are in columns 5 and~6, and the corresponding
masses $M_{\rm gas}$ and~$M_{\rm DM}$ in columns 7 and~8.
Columns 9, 10, and~11 show the number of gas particles $N_\disk$,
the corresponding baryonic mass~$M_\disk$, and the ratio $M_\disk
/\Omega_bM_{200}$ for the cold disk present at the end of each
simulation.

The time evolution of $j_{disk}$ and $M_{disk}$
for the WDM2 MR simulations with $\Omega_b$=0.05 of the forming
disk galaxies S1-S4 is shown in Figures 5-8 together with those of the
corresponding ``passive'' CDM simulations (from SLGV99) and of the WDM2
HR simulations of galaxies S1-S4 (though not $j_{disk}$ for galaxy S1 - see
below). 
The WDM2 simulations are clearly
producing disks with considerably larger $j_{disk}$ than the CDM simulations,
whereas the growth of disk mass $M_{disk}$ is fairly similar for the two
types of dark matter. Comparing the WDM2 MR and HR simulations qualitative
agreement is found in the $j_{disk}$ and $M_{disk}$ evolution between the
two types of simulations. This is gratifying, since the substructure
in the forming galaxies with this kind of dark matter in general is resolved 
even in the MR simulations (with one exception, as discussed below).
For any given galaxy, however, $j_{disk}(t)$ for the HR simulations generally
exceeds $j_{disk}$ for the MR simulations. This is particularly evident
towards the end of the simulations ($z \sim$ 0), where the ratio of the two
typically is 1.5-2.5. This strongly suggests that part of the angular momentum
problem is due to numerical effects: We have performed \TSPH simulations of
(idealized) smooth, disk galaxy forming cooling-flows, in differential rotation
in a rigid, spherical halo potential to study the effects of spurious angular
momentum transport related to SPH. From these test simulations we find that
increased resolution leads to less angular momentum transport and resulting
central, cold disks of higher specific angular momentum and that
the angular momentum transport is mainly due to the shock capturing, artificial
viscosity used in SPH. We also find that
going from the standard Monaghan-Gingold viscosity (Monaghan \& Gingold
\cite{MG83}) used in this work to the shear-free Balsara (\cite{B95}) viscosity
does not result in higher specific angular momentum disks at a given 
level of resolution. These issues will be discussed in a future paper - see 
also the brief discussion in SLGV99.

Galaxy S1 is very different from the others: At about $t \sim$ 4 Gyrs
the central galaxy develops into a kinematically complex system through 
accretion of counterrotating hot gas and a cold gas merging event. By then
it consists of
an inner, high density disk surrounded by an outer, more extended but lower
density counterrotating disk. In the MR simulations the spin vector of the 
inner disk flips around and aligns itself with that of the outer disk during
the next 2-4 Gyrs. In contrast, the inner disk in the HR simulation does not
``flip'' relative to the outer during the remaining $\sim$ 9 Gyrs of the 
simulation. This clearly indicates that the gasdynamical (and perhaps
also gravitational) resolution is not sufficient to resolve the long-lived,
discontinuous kinematics. We chose to plot and give values for the specific
angular momenta of the final disks formed in the MR simulations of galaxy
S1, as these disks are kinematically relaxed. But given the possible
resolution problems in these simulations, we do not use the results in any
of the following quantitative estimates of the angular momentum properties of 
disks formed in
WDM simulations. We do not give specific angular momenta results for the HR
simulation of galaxy S1, as this system is kinematically unrelaxed during
most of the simulation, including the final state. Given the observational
variety of galaxies, including some with counterrotating disks (e.g.,
Thakar \& Ryden \cite{TR98} and references therein), we find it
almost reassuring that not {\it all} simulated galaxies end up as extended, 
kinematically coherent disks, as long as most do!

The time evolution of $j_{disk}$ and $M_{disk}$ of galaxies S1-S4
for the WDM2 MR simulations with $\Omega_b$=0.05 and no UVX background
radiation field is shown in Figure 9 (runs \#11-14). 
Not surprisingly,
the final disk masses are somewhat larger (on average by about 25\%) than
for the corresponding runs with a UVX field. This is due to the
somewhat more efficient radiative cooling and lack of photo-heating in the 
absence of a UVX field. 
Moreover, the specific angular momenta of the final disks are on average
about 45\% larger than for the runs with a UVX field - we comment on this
below.

Finally, the time evolution of $j_{disk}$ and $M_{disk}$ of galaxies S1-S4
for the WDM2 MR simulations with $\Omega_b$=0.10 and a UVX field is shown in 
Figure 10 (runs \#15-18). The final disk masses are about a factor of 2.5 
times those of the corresponding $\Omega_b$=0.05 runs. This is not far
from the factor of about $2^{3/2}$ expected from cooling flow theory assuming 
an isothermal dark matter potential (Sommer-Larsen \cite{SL91}). The
specific angular momenta of the final disks are about a factor 2.8 times
those of the corresponding $\Omega_b$=0.05 runs. This is significantly more
than the factor of about$2^{1/2}$ expected from cooling flow theory
(Sommer-Larsen \cite{SL91}) - we also comment on this below.  

In order to compare the masses and specific angular momenta of the final disks
to observations a characteristic circular velocity, $V_c$, is assigned to the
model disk galaxies in the following way: Courteau (\cite{C97}) compared for 
Tully-Fisher applications optical rotation curves with 21 cm linewidths of disk
galaxies. He found from the data that the circular 
velocity, $V_{2.2}$, at 2.2 exponential disk scale-lengths from the center 
(where the circular
velocity of the disk alone peaks) gives the best match to the 21 cm 
linewidths. Since the disks of our model galaxies are not perfectly 
exponential and generally still somewhat too concentrated compared to
observations we determine $V_c$ = $V_{2.2}$ iteratively. Given the dark
matter part of the rotation curve and the disk mass, both taken from the
simulations, and using the observational relation between the median
exponential disk scale-length and the characteristic circular velocity
\begin{equation}
\bar{R}_d = 1.9 ~h^{-1} ~\left(\frac{V_c}{200 ~\rm{km/s}}\right)^{1.05} 
~~\rm{kpc}
\end{equation}
(SLGV99), $V_c$ = $V_{2.2}$ is determined in the following way:
Starting from an initially adopted $V_{2.2}$ (= $V_c$) the corresponding 
characteristic
scale-length is determined using (14). Assuming that the ``real'' disk is
exponential and given the disk mass $M_{disk}$ and scale-length $R_d$ the disk
part of the rotation curve is determined. This is then added in quadrature 
to the dark matter part of the rotation curve and finally $V_{2.2}$ is 
determined from the combined rotation curve. This value is then used as input
to the above procedure, which is repeated until the input $V_{2.2}$ equals
the output $V_{2.2}$.

As the baryonic mass is somewhat more centrally concentrated than the
exponential disk of the same mass the dark matter halo may be somewhat too
``pinched''. This means that $V_c$ = $V_{2.2}$ determined in this way 
actually is an upper limit in this connection. We estimate, however, that the 
effect of the
extra pinching is quite small: As an example we show in Figure 11 the dark
matter part of the rotation curve $V_{c,DM}(R)$ (upper panel) and 
$M_{disk}(R)$ (lower panel) for the four simulations runs \#4, 7, 11 and 15 
of galaxy S1 at
$z$=0. As can be seen from the Figure, the final dark matter halo rotation
curves are fairly insensitive to the quite different amounts of baryonic mass
inside of $R_{2.2} \simeq$ 10-15 kpc for $h$=0.5. Another possible concern
is that part of the cooled-out mass may end up in a central component/bulge,
which may change the estimate of $V_{2.2}$. Assuming that no more than 20\%
of the baryonic mass is in such a component for Sb-Sd galaxies (see section 
5.2) we find, however, that this leads to an average increase of $V_{2.2}$ of 
no more than about 1\%, so we neglect this effect as well. Finally, for the
arguments given in this paper, we can neglect the effects of a possible 
truncation of the disk at $\sim 4~R_D$ (van der Kruit \& Searle \cite{KS82}).

The specific angular momenta $j_{disk}$ and iteratively determined
characteristic circular velocities $V_c$ = $V_{2.2}$ of the final disks
in the 18 simulations are given in Table 3. The Table also lists
our computed values for the
dimensionless spin parameter $\lambda \equiv J|E|^{1/2}/GM^{5/2}$
(column~3) evaluated at the {\em infall radius} $r_{inf}$
(column~5). This radius is defined by $M_{DM}(r_{inf})/(1-\Omega_b) =
M_{disk}/\Omega_b$, and is of order 120-200~kpc in all our runs.
It represents a characteristic radius, at the present time, of the
dark matter originally associated with the amount of gas currently in
the disk.
In Figure 12 we show the ``normalized'' specific angular momenta 
$\tilde{j}_{disk} = j_{disk}/V_c^2$ of the final disk galaxies formed in all
WDM2 runs (except run \#7)
as a function of $V_c$. As argued by SLGV99 one expects $\tilde{j}_{disk}$ to 
be almost independent of $V_c$ on both theoretical
and observational grounds. Also shown in the figure is the median 
``observed'' value of $\tilde{j}_{disk}$, calculated using equation (14) and 
\begin{equation}
\tilde{j} = \frac{j}{V_c^2} = \frac{1.68 R_d}{V_c} 
\end{equation}
(SLGV99), and the observational 1-$\sigma$ and 2-$\sigma$ contours.

For the MR WDM2 simulations of galaxies S2-S4 with $\Omega_b$ = 0.05 and
including effects of a UVX field the median $j \simeq$ 500 kpc~km/s and
the median $\tilde{j}_{disk}$ is
0.61 $\pm$ 0.11 dex below the ``observed'' value, corresponding to a factor 
of 4.
This is almost an order of magnitude better than what is obtained with
``passive'' CDM simulations - see, e.g., Navarro \& Steinmetz (\cite{NS97})
and SLGV99.
 
For the HR WDM2 simulations of galaxies S2-S4 with $\Omega_b$ = 0.05 and
including effects of a UVX field the median $j \simeq$ 940 kpc~km/s and
the median $\tilde{j}_{disk}$ is (just)
0.32 $\pm$ 0.07 dex below the observed, corresponding to a factor of two.
As noted above this strongly suggests that part of the angular momentum
problem is related to numerical problems. 

For the MR WDM2 simulations of galaxies S2-S4 with $\Omega_b$ = 0.05 and
no UVX field the median $j \simeq$ 730 kpc~km/s and the 
median $\tilde{j}_{disk}$ is
0.45 $\pm$ 0.13 dex below the observed, corresponding to a factor of 2.8.
We interpret
the larger specific angular momenta obtained for the final disks relative
to the simulations including a UVX field as being due to the combination
of two effects: a) The more efficient
cooling results in more high angular momentum gas from the outer regions of
the halo to be deposited onto the disk during the simulations - as discussed by
Navarro \& Steinmetz (\cite{NS97}) and b) This increased
cooling efficiency and lack of photo-heating does {\it not} lead to the
cool-out of a very large number of gas clouds in small mass dark matter halos 
($V_c \la$ 40 km/s, Efstathiou \cite{Ef92}, Quinn \etal \cite{Q.96}) in our 
WDM2 simulations compared to CDM, because the
number of such small dark matter halos is greatly reduced for WDM2 relative
to CDM. This could otherwise lead to an increased angular momentum
transfer from the gas clouds to the main dark matter halo by dynamical
friction, as discussed in section 1.

The first of the two effects discussed above is even more pronounced
in the MR WDM2 simulations with $\Omega_b$=0.10 and a UVX field due to the
$\sim 4$ times higher cooling rates compared to the similar simulations with
$\Omega_b$=0.05: For galaxies S2-S4 the median $j \simeq$ 1380 kpc~km/s and
the median $\tilde{j}_{disk}$ is
(just) 0.36$\pm$0.12 dex below the observed corresponding to a factor of 2.3.

Summarizing, we find that increasing the mass resolution by a factor of 4
(8 for DM) leads to the formation of disk galaxies with final specific 
angular momenta, which are almost a factor two larger than those of the
otherwise identical medium resolution simulations and just a factor of about 
two smaller
than those of observed disk galaxies. Moreover, the median value of
the spin parameter $\lambda$ for the three HR simulation galaxies 
S2-S4 is 0.039.
Assuming a theoretical median value of $\lambda$ of 0.05 (Barnes \&
Efstathiou \cite{BE87}; Heavens \& Peacock \cite{HP88}) the discrepancy is 
further reduced to a factor of 1.6.
Given that the rather modest increase in mass resolution
reduces the discrepancy by almost a factor of two, we find it
plausible that the remaining factor of $\sim 1.6$ may be gained by
increasing the mass resolution even further, but this obviously has
to be checked with very high resolution simulation.
Furthermore we also find that going from a baryonic fraction of $f_b$=0.05
to $f_b$=0.10 (which may well be more realistic) also improves the
situation considerably for the MR simulations, as discussed above. To
illustrate this we show in Figure 13 a face-on view and in Figure 14
an edge-on view of the final galaxy formed in run \# 18 (this has the
largest specific angular momentum of all the galaxies formed, $j_{disk}
\simeq$ 2000 kpc~km/s). Further, in Figure 15 we show the (azimuthally 
averaged) surface
density profiles of the final disks in runs \#15-18. Clearly, it is no longer
a problem to form extended disks in cosmological simulations!

All in all we find that 
the disk galaxy angular momentum problem can be significantly alleviated
(and we speculate: perhaps even completely resolved)
by going to the WDM cosmological structure
formation scenario, provided that the WDM free-streaming mass is 
$M_{f,WDM} \sim 10^{11} h^{-1}\Msun$ within about a factor of three if
$\Omega_M$=1. We shall touch briefly on the $\Omega_M < 1$ case in
section~\ref{s:conclusions}. 

\subsection{The Tully-Fisher relation and mass-to-light ratios of disk 
galaxies}

In Figure 16 we show $M_{disk}(V_c)$ of the final disk galaxies formed in 
all 16 WDM2 runs together with the $I$-band Tully-Fisher
relation (TF) of Giovanelli \etal (\cite{G.97}) for $h$=0.5, converted
to mass assuming mass-to-light ratios $(M/L_I)$ = 0.25, 0.5 and 1.0
in solar units (used throughout). The
slope of the ``theoretical'' TF matches that of the observed very well
for a constant mass-to-light ratio, and for $h$=0.5 the required $(M/L_I) 
\simeq$ 0.6. This value is in fair agreement with some estimates of the
$(M/L_I)$ of disk galaxies in the literature and somewhat low relative
to others:
Persic \& Salucci (\cite{PS92}) find a dynamical estimate of $(M/L_B)$ = 
1.24$h$ for disks, corresponding to $(M/L_B)$ = 0.62 for
$h$=0.5 or $(M/L_I) \simeq$ 0.4 assuming Sbc type colours (see also
Persic \etal \cite{P.96}). Syer \etal
(\cite{Sy.97}) find $(M/L_I) < 1.9h$ from disk stability arguments
corresponding to $(M/L_I) <$ 0.95 for $h$=0.5. The $(M/L)$s obtained from
stellar population synthesis depend strongly on the assumed initial mass
function (IMF), in particular the effective lower mass cut. Recent
determinations of the $(M/L)$ of disk galaxies based on what is thought to be 
realistic IMFs range from
$(M/L_B) \simeq$ 1.9 (Fukugita \etal \cite{F.98}), 
corresponding to $(M/L_I) \simeq$ 1.3, to $(M/L_I) \simeq$ 1.0 for Milky Way 
sized disk galaxies and slightly less for smaller galaxies (Boissier \&
Prantzos \cite{BP99}).
Adopting (more realistically) $h$=0.7 increases the dynamical estimates to
$(M/L_I) \simeq$ 0.6 (Persic \& Salucci) and $(M/L_I) <$ 1.3 (Syer \etal),
bringing them in better agreement with the stellar population synthesis ones.
One can show
that the final virial mass and final cooled-out disk mass will scale 
approximately as $h^{-1}$ for the simulations, everything else being
unchanged. As the inferred absolute luminosity of the disk galaxies scales
as $h^{-2}$, one would expect $(M/L_I)$ to increase by about 40\% by going
from $h$=0.5 to $h$=0.7. However, since the characteristic length-scale of the
CDM 
(and hence WDM) power-spectrum scales as $h^{-2}$ (for fixed $\Omega_M$)
and not  $h^{-1}$ the magnitude of linear density fluctuations
on galactic mass scales is approximately 30\% larger for $h$=0.7 than
for $h$=0.5, given the same value of $\sigma_8$. To test the effect of this
we carry out four MR WDM2 simulations (with $\Omega_b$=0.05 and including the 
UVX field) of the formation of disk galaxies S1-S4 with $h$=0.7, but otherwise
similar to runs \#2 and \#4-6 (in particular with $\sigma_8$=0.5, as before).
For the resulting 4 galaxies we find $(M/L_I) = 0.62\pm0.02$ compared to
$(M/L_I) = 0.54\pm0.02$ for the similar $h$=0.5 runs, an increase of (only)
11$\pm$5\%. Assuming this modest increase to be generally applicable, we
would expect a $(M/L_I) \simeq$ 0.7 for WDM2 simulations with $h$=0.7
($\Omega_M=1, \Omega_{\Lambda}=0, \sigma_8=0.5$).   

All in all, the result of our WDM2 simulations suggest that values of 
$(M/L_I) \simeq$0.6-0.7 for $h$=0.5-0.7 are required to match the observed
Tully-Fisher relation. Such values of $(M/L_I)$ for disks are fairly consistent
with dynamical estimates, but somewhat low compared to stellar population
synthesis values.

Steinmetz \& Navarro (\cite{SN99}) and Navarro \& Steinmetz (\cite{NS99})
carried out {\sc GRAPESPH}, SCDM and $\Lambda$CDM simulations of disk galaxy 
formation. They found that a $(M/L_I) \simeq$ 0.5 is required to match the
results of the simulations to the observed $I$-band Tully-Fisher relation. 
Their required $(M/L_I)$ value is comparable to, though somewhat lower than
ours. Since this difference could be related to the use of different types
of dark matter, we determined the $(M/L_I)$ required to match the results
of our CDM simulations corresponding to runs \#2 and \#4-6 (described in
SLGV99) to the observed TF relation. We find a $(M/L_I) = 0.51\pm0.04$ for
our CDM simulations compared to $(M/L_I) = 0.54\pm0.02$ for the corresponding
WDM2 simulations ($h$=0.5), so there is no indication that going from CDM
to WDM leads to larger $(M/L_I)$s. It is not clear what leads to the (small)
difference between the results of the above authors and ours, but it may be
related to the use of different normalizations, $\sigma_8$, of the power
spectra.

We now estimate a general upper limit to
the $(M/L_I)$ of disk galaxies by assuming that disks are ``maximal'', i.e.
as dominant in the inner part of the galaxies as possible given the rotation
curve constraints: Maximal disk galaxies have $\gamma \equiv V_{2.2,D}/
V_{2.2} \simeq 0.85$ (e.g., Sackett \cite{S97}), where $V_{2.2,D}$ is the
(peak) rotation velocity of the disk alone at $R_{2.2}$. 
The disk mass can be expressed
as $M_{D} = f_{D}M_b$, where $M_b$ is the total,
baryonic disk galaxy mass (central component + bulge + disk - neglecting the
stellar halo and, for notation, assuming that dark matter is non-baryonic). 
For an exponential disk
\begin{equation}
V_{2.2,D} = 0.62 ~\sqrt{\frac{GM_{D}}{R_D}} ~~.
\end{equation}
Using this, equation (14) and the definition of $\gamma$ we then obtain 
\begin{equation}
M_b = 4.2\cdot 10^{10} ~h^{-1} 
~\left(\frac{f_D}{0.8}\right)^{-1} \left(\frac
{\gamma}{0.85}\right)^2 \left(\frac{V_c}{200 ~\rm{km/s}}\right)^{3.05} 
~\Msun ~~.
\end{equation}
Following Giovanelli \etal (\cite{G.97}) the $I$-band TF relation can be 
expressed as
\begin{equation}
L_I = 2.1\cdot 10^{10} ~h^{-2} ~\left(\frac{V_c}{200 ~\rm{km/s}}\right)
^{3.07} ~L_{I,\odot} ~~,
\end{equation}
so the $I$-band mass-to-light ratio of a disk galaxy is given by
\begin{equation}
(M/L_I) = 2.0 ~h
~\left(\frac{f_D}{0.8}\right)^{-1} \left(\frac
{\gamma}{0.85}\right)^2 \left(\frac{V_c}{200 ~\rm{km/s}}\right)^{-0.02} 
~\Msun ~~.
\end{equation}
Adopting $f_D \ga 0.8$ for Sb-Sd galaxies (e.g., Broeils \& Courteau \cite
{BC97}), $\gamma \la 0.85$ (maximal or sub-maximal disks) and neglecting
the very weak $V_c$ dependence, we obtain
\begin{equation}
(M/L_I) \la 2.0 h
\end{equation}
for intermediate to late type disk galaxies, which for $h$=0.7 translates
into $(M/L_I) \la 1.4$. This is likely to be a generous upper limit as
disk galaxies may well be sub-maximal ($\gamma < 0.85$) - see, e.g.,
Courteau \& Rix (\cite{CR99}). The estimated upper limit in equation (20) 
is consistent with
the various dynamical and spectrophotometric estimates discussed above, for 
$h \sim 0.7$.

\subsection{The total baryonic mass of the Milky Way as a constraint on
disk galaxy formation simulations}

Another observational check of the results is obtained by comparing the
final disk masses as a function of $V_c$ with the observed, baryonic mass of
the Milky Way: Flynn \& Fuchs (\cite{FF94}) find a dynamically determined
local disk surface density of 52$\pm$13 \Msun/pc$^2$, out of which about
20-25\% is gas. Assuming that the Galactic disk is exponential and that the
Sun is situated at $R$=8 kpc, the total disk mass can be calculated as a
function of the exponential scale-length $R_D$. The result is shown in
Figure 18 (solid line) together with the observational 1-$\sigma$ contours.
This provides a lower limit to the baryonic mass of the Milky Way (the 
radial gas surface density profile is somewhat shallower than that of the 
stars, but the effect of this on the total disk mass is small, especially
since only about 10\% of the baryonic mass of the Milky Way is gas). 

An upper
limit can be obtained by taking into account the mass of the central component
(CC)/bulge also. This mass component is highly centrally concentrated, so
for our purpose we simply assume that it is located at the center and
characterized by a Keplerian rotation curve. We now demand that
the combined disk/CC/bulge rotation curve should not exceed the observed
value of $\la$ 200 km/s at $R$=3 kpc (Rholfs \etal \cite{R.86}) and in this
way obtain the CC/bulge mass upper limit shown in Figure 18 (short-dashed line)
together with lower and upper 1-$\sigma$ contours (long-dashed lines) 
corresponding to the upper and lower ones for the disk mass.
The combined disk/CC/bulge upper
1-$\sigma$
mass limit for the Milky Way is shown as a heavy solid line in the Figure.
We conclude that the baryonic mass of the Milky Way is somewhere in the
range 3 - 7 $\times$ 10$^{10}$ \Msun with a likely value of about 
5 $\times$ 10$^{10} \Msun$.
We adopt a characteristic circular
velocity of the Galaxy of $V_c$ = 220 km/s with a 10\% uncertainty
and plot the result in Figure 16. Clearly, our
results are compatible with the Milky Way baryonic mass constraint for
$h$=0.5 (as well as $h$=0.7 - see below), given the uncertainties.

Our determination of the baryonic mass of the Milky Way provides an
effectively independent way of estimating the $I$-band mass-to-light ratio
of a disk galaxy: Assuming $M_{b,MW}=5\pm2\times 10^{10} \Msun$, 
$V_{c,MW}=220\pm20$ km/s, $h=0.7\pm0.1$ and that the $I$-band luminosity of
the Milky Way is given by equation (18) with a 40\% uncertainty (approximately
the dispersion in luminosity at a given $V_c$ of the TF relation), and
adding all uncertainties in quadrature yields $(M/L_I)_{MW}$=0.87$\pm$0.57.
This further strengthens the case for a low value of the typical $(M/L_I)$ 
of disk galaxies ($(M/L_I) \la 1$).

\section{Physical candidates for warm dark matter.}
\label{s:candidates}

Warm dark matter is usually considered as exotic in
comparison with ``normal'' cold dark matter. However 
warm dark matter particles (we will
call them {\it warmons} in what follows) are by no means less
natural than cold dark matter ones. Of course naturalness is
a rather vaguely defined notion and we assume that some particles
are natural candidates for dark matter if their existence is
predicted or requested by the theory independently from dark matter
problem. The most popular candidates for cold dark matter are either
lightest supersymmetric particles with the mass of a few hundred
GeV which would be stable if the so called R-parity is conserved, 
or very light axions with the mass about $10^{-5}$ eV. 
There are several equally natural candidates for
warm dark matter particles as well. Their properties are fixed by
the condition that they should be stable on the cosmological time
scale (the life-time should be larger than the universe age, 
$\tau_w > t_U \approx 10^{10}$ Gyr) and their mass should be 
sufficiently light $m_w \sim 1$ keV, to give a
free-streaming mass of $\sim 10^{11} \Msun$, cf. equation (6) (in this
section we use the system of units with $c=\hbar=k_B=1$).
The simplest possibilities for the warmons are either massive 
neutrinos or pseudo-Goldstone bosons. 

It is natural to start with massive neutrinos, firstly, because neutrinos,
in contrast to other possible dark matter particles are known to exist 
and, secondly, they may be massive.
However the interpretation of atmospheric neutrino anomaly by oscillations
between $\num$ and $\nut$ or a sterile neutrino $\nu_s$
demands a very small mass difference of oscillating neutrinos,
$\delta m^2 \approx 0.01\,\, {\rm eV}^2$ (Super-Kamiokande Collaboration
\cite{sk}). An explanation of the solar neutrino deficit by neutrino 
oscillations also demands a very small mass difference, now between 
$\nu_e$ and some other active or sterile neutrinos 
(see, e.g., Bahcall \etal \cite{bahcall98}). 
Other neutrinos may have masses in the keV range and be warmons:
An example of warm dark matter created by sterile neutrinos 
was recently considered in Dolgov \& Hansen \cite{dolgov00}.

According to the 
Gerstein-Zeldovich (\cite{gz}) and Cowsik \& MacLelland (\cite{CM72}) 
neutrinos with the {\it normal weak 
interactions} have the cosmological mass density,
$ \Omega_\nu  h^2 =  {m_\nu /94 {\rm eV}}$
so keV neutrinos would over-close the universe.
A possible  way to save neutrinos as warmons is to assume 
that they have a new stronger interaction, diminishing
their number density by one-two orders of magnitude
Correspondingly the cross-section of their annihilation should be
\be
\sigma_{ann} \approx {10 \over 94\, {\rm eV}\, m_{Pl} \Omega_w h^2}
 \approx 4\cdot 10^{-39} {\rm cm}^2 /(\Omega_w h^2 ) ~~.
\label{sigmaw}
\ee

Due to this interaction the neutrinos would decouple at 
$T \approx m_\nu/5$.
Until then their mean-free path is determined by the
annihilation and would be considerably shorter than the
horizon. This could reduce the scale at which perturbations are erased
in comparison with more conventional dark matter, which is free-streaming
for $T \ga m_\nu$. However in this model
there exists a new mechanism for dissipation of perturbations at small
scales, namely $\nu\nu$-annihilation. The larger the number density
of neutrinos the stronger the annihilation 
and thus neutrinos disappear faster in regions with higher neutrino 
density, reducing the density contrast. 

Elastic scattering of neutrinos (or other warmons,
e.g. light scalars) may be stronger than the annihilation and
the particle mean free path may be much smaller
than galactic scales. This could lead to a lowering of the central 
concentration of dark matter halos - 
see, e.g., Spergel \& Steinhardt (\cite{spergel99}). 

Another warmon model can be realized with early decoupled
keV particles whose number density is diluted by a subsequent entropy 
release. Hence the present day ratio of the number
densities of warmons and CMB photons would be:
\be
(n_w /n_\gamma)_0 = 43 g_w /11 g_{dec} ~~,
\label{nwngamma}
\ee
where $g_{dec}$ is the effective number of particle degrees of freedom
at warmon decoupling and $g_w = (n_w /n_\gamma)_{dec}\approx 1$.
The contribution of such warmons to the present day ``mass density'' 
$\Omega h^2$
would be
\be
\Omega_w h^2 = 1.5\cdot 10^2~ m_w (g_w/g_{dec})
\label{omegaw}
\ee
where the warmon mass is expressed in keV. This expression is
very similar to equation (4) in section 2 for $g_w\sim$1.
In the minimal standard model the number of particle species above the 
electroweak scale is $g_{dec}$= 107.75 so the suppression factor is too small
to get a small $\Omega_w h^2 \sim 0.1$ with keV mass warmons.
Only for very weak interactions and if the rank of the unification 
group is sufficiently high the entropy suppression factor would 
be large enough to permit keV warmons with
$\Omega_w h^2 \sim$ 0.1. Another mechanism of a large entropy release 
could be cosmological first order phase transitions.

In the case of earlier decoupled particles their 
momentum distribution would be considerably softer 
so they would have a smaller free-streaming scale.
The ratio of the warmon temperature,
when they are relativistic, to the photon
temperature can be expressed as: 
\be
{T_w \over T_\gamma} = \left( {43 \over 11 g_d}\right)^{1/3} 
= 0.3 \left({ \Omega_w h^2 \over g_w m_w }\right)^{1/3} ~~.
\label{twtgamma}
\ee
The physical warmon path in the expanding reference frame is:
\be
l_w = 2.6\cdot 10^6\,{\rm sec}\,m_w^{-8/3}(\Omega_w h^2/g_w)^{2/3}~~. 
\label{lw}
\ee
The corresponding comoving free-streaming scale $R_{f,w}$
is given by $(1+z_w) l_w$, where
\be
(1+z_w) = \frac{T_\gamma}{T_\gamma^{(0)}} = \frac{m_w^{4/3}}{0.9 
T_\gamma^{(0)}}~\left(\frac{g_w}{\Omega_w h^2}\right)^{1/3} =
4.7\cdot 10^{6}~m_w^{4/3}~\left(\frac{g_w}{\Omega_w h^2}\right)^{1/3}~~,
\label{opz}
\ee
and $T_\gamma^{(0)}$=2.4$\cdot 10^{-7}$ keV (corresponding to 2.73 K)
is the present temperature of the CMB. Hence:
\be
R_{f,w} = (1+z_w) l_w =
0.23~(g_{dec}/100)^{-4/3}~(\Omega_w h^2)^{-1}
~g_w ~~{\rm Mpc} ~~,
\label{rfwdm}
\ee
very similar to equation (3) in section 2 for $g_w\sim$1.

A generic theoretical problem is the required smallness of the warmon mass. 
We know that, though do not understand why, most ``normal'' 
neutrinos probably are {\it too} light, so
to be warmons, they must 
either possess stronger than normal 
weak interactions or vice versa, be almost sterile.  A possible example
of the latter are the
right-handed neutrinos considered by Malaney \etal (\cite{malaney95}) and
Colombi \etal (\cite{colombi96}).

If neutrinos are not WDM particles, then the next best candidate
is probably the majoron (Berezinsky \& Valle \cite{berezinsky93}, Babu \etal
\cite{babu93}, Dolgov \etal \cite{dolgov95}). 
Majorons could quite naturally have a mass in the keV 
range and could give
$\Omega_w h^2 \sim 0.1$. If majorons indeed exist, 
they could either be warmons or would open such a possibility 
to neutrinos because their coupling to $\bar \nu \nu$
introduces a new interaction that would permit 
to avoid the Gerstein-Zeldovich limit.

There is another possible way of cosmological creation of pseudo-Goldstone 
bosons, through a phase transition. 
In this case the particles are produced at rest
and a mechanism to warm them up is 
required. This could be realized if they have a sufficiently strong 
self-interaction, $\lambda J^4$ (where $\lambda$ is the coupling constant
and $J$ is the majoron field). 
Majorons would predominantly disappear in 
$4J\rar 2J$-reactions. Thus this process not only reduces the 
number density of majorons but also heats them up. The freezing of
species by $4 \rar 2$ or $3\rar 2$-reactions and properties of the
corresponding dark matter has been considered by 
Dolgov (\cite{dolgov80}), Carlson \etal (\cite{carlson92}), Machacek \etal
(\cite{machacek94}), de Laix (\cite{laix95}) and Dolgov \etal 
(\cite{dolgov95}). 
Depending upon the efficiency of the cooling and on the initial
majoron momentum distribution this model would supply dark matter 
somewhat warmer or colder than warm. 

It is possible that both thermal and phase transition mechanisms are 
effective, so that the same particles form both cold and 
warm dark matter. Thus,
compared to CDM, one could have a smaller, but {\it non-vanishing} 
power on mass
scales less than the WDM free-streaming scale ($\sim 10^{10}$-$10^{11}
M_{\odot}$). 
Another case of interest is that some neutrino species are 
heavier than majorons and decay into majorons and 
lighter neutrinos. This was considered in several papers dealing with 
unstable dark matter - see, e.g., Dolgov (\cite{dolgov99}) and references
therein. One may also keep neutrinos rather heavy (in the MeV range), 
unstable and relatively short-lived, while keV majorons, produced by neutrino 
decays when the number density of neutrinos
was already Boltzmann suppressed, would be dark matter particles. 
According to e.g.  Dolgov \etal \cite{dolgov97}
a conflict with big bang nucleosynthesis can be avoided.
A new zoo of interesting candidates for dark matter particles 
is opened up if mirror or shadow worlds exist. 
A list of references and a discussion of the subject can be
found, e.g., in Dolgov (\cite{dolgov99}). Shadow world neutrinos, as 
considered by 
Berezhiani \& Mohapatra (\cite{berezhiani95}) and Berezhiani \etal 
(\cite{berezhiani96}) would look in our world as sterile ones and
might have a keV mass.
Finally, new dark matter particles, that might be warmons,
have been recently considered by Garriga \& Tanaka 
(\cite{multidim}), Arkani-Hamed \etal (\cite{multidim1}) and Cs{\'a}ki \etal
(\cite{multidim2}) in the framework of 
multidimensional cosmologies. Phenomenologically they are rather similar
to shadow particles.

In summary, there are plenty of particle physics candidates for
warmons. Their properties and initial momentum distributions may be 
significantly different. They could be self-interacting or sterile, long-lived
or absolutely stable. To mention a few, they could be neutrinos with weaker
or stronger couplings than normal, majorons (light pseudo-Goldstone bosons)
or shadow or mirror world neutrinos.

\section{Discussion and conclusions}
\label{s:conclusions}

In conclusion we find that the disk galaxy angular momentum 
problem can be considerably alleviated (and we speculate: perhaps even 
completely resolved) by going from the CDM to the WDM cosmological 
structure formation scenario without having to resort to star-burst driven
feedback processes at all. The main reason for this is that the inflow of
gas onto the forming disk is more smooth and coherent for WDM than for
CDM, resulting in considerably larger disk angular momenta, as discussed
by SLGV99.

For $\Omega_M$=1 our results suggest that this 
requires the WDM free-streaming mass to be $M_{f,WDM} \sim 10^{11} h^{-1}
\Msun$ within about a factor of three, indicating that extreme fine-tuning
is not required. This range of characteristic free-streaming masses corresponds
to characteristic wavenumbers $k_c \sim$ 5-10 $h$~Mpc$^{-1}$. For wavenumbers
about 2-3 times larger than this the WDM power spectrum is significantly 
reduced relative to the CDM one, cf.~Figure 1. 
In the interesting work by Croft \etal (\cite{C.99}) the 
linear power spectrum is determined ``observationally'' from the Ly$\alpha$ 
forest in spectra
of quasars at $<$z$>$=2.5 for wavenumbers in the range
0.3-0.5$\la k \la$ 2-3 $h$~Mpc$^{-1}$. In this range of wavenumbers the 
measured $P(k)$ is found to compare well with CDM power spectra. If this
method can be pushed to about 4 times larger wavenumbers, it would be possible
to test directly the hypothesis that dark matter is warm, with the 
characteristics proposed in this work, rather than cold. In fact, McDonald
\etal (\cite{Mc.99}) have been able to push the method to about twice as
large wavenumbers by using high dispersion spectra of quasars at somewhat
larger redshifts (see also Nusser \& Haehnelt \cite{NH99}), but this is
probably as far as one can get using the Ly$\alpha$ forest method (R. Croft,
private communication). Another, but much more indirect way of obtaining
information about power on small scales may come from the comparison of
observations and models of damped Ly$\alpha$ systems - see, e.g., Haehnelt
\etal (\cite{H.98}) and McDonald \& Miralda-Escud{\'e} (\cite{MM99}).

Currently there are many indications that $\Omega_M$ probably is less than 
unity, 
$\Omega_M \simeq 0.3$, say. For such OWDM as well as $\Lambda$WDM models
the cluster normalized value of $\sigma_8$ will be $\simeq 1$, rather than
0.5 for the $\Omega_M$=1 considered in this paper. Also, the late linear
growth is slower in the low $\Omega$ models. Taking this together implies
that the density fluctuations are about 1.5-2 times larger at redshifts $z \ga 
\Omega_M^{-1}-1$ and consequently that galaxy formation is somewhat faster and 
more intense in low $\Omega_M$ models. For such cosmologies it therefore seems 
reasonable to assume that the WDM free-streaming mass-scales, which will allow 
large, high specific angular momentum disks to form, will be comparable to, 
but no less
than the ones found in this work for $\Omega_M$=1. From this constraint and
equation (6) it follows that the mass of the warm dark matter particle(s) has 
to be of the order 1 keV with a fairly weak dependence on $\Omega_{WDM}$. 

We find a slope of our ``theoretical'' Tully-Fisher relation which matches
that of the observed $I$-band TF relation very well. In terms of the 
normalization of the TF relation we find agreement provided
that the mass-to-light ratio of disk galaxies is $(M/L_I) \simeq 0.6-0.7$ for 
$h$=0.5-0.7. We argue that this is in reasonable agreement with various recent
estimates of $(M/L_I)$, including two given in section 5.2 of this paper.
The discrepancy of a factor of $\sim4-5$ in absolute luminosity found by
Steinmetz \& Navarro (\cite{SN99}) and Navarro \& Steinmetz
(\cite{NS99}) in matching the observed TF relation, using similar kinds of
CDM disk galaxy formation simulations, is at least partly caused by the 
adoption of a larger mass-to-light ratio ($(M/L_I) \sim 2$) by these authors.
We find no indication that going from CDM to WDM increases the $I$-band
mass-to-light ratio required for the model disk galaxies to match the
observed TF relation. This is agreement with the findings of Moore \etal
(\cite{M.99b}) that a suppression of power on small scales similar to the
one discussed in this paper does not change the structure of the resulting
dark matter halos. However, in their simulations as well as ours the 
possibility of having WDM particles, which initially have a non-zero
velocity dispersion (due to free-streaming) was not taken into account. 

A potential problem for the
WDM scenario might be a lack of a sufficient rate of early galaxy
formation to match observations: Lyman break galaxies are routinely
found at redshifts 3-4 and galaxies have been detected at even higher 
redshifts, with a possible current record of $z$=6.7 (Chen, Lanzetta \&
Pascarelle \cite{CLP99}).
Assuming that galaxies form from $n$-$\sigma$ peaks, where $n \sim$ 2-3 (e.g., 
Ryden \& Gunn \cite{RG87}), and using linear theory
it follows that the typical formation redshift of galaxies of mass $M$ will
be $z_f \simeq n\sigma(M;z=0)/\delta_c - 1$,
where $\delta_c$=1.69 (see White \cite{W93}).
From Figure 2 it follows that
for the characteristic WDM free-streaming masses relevant for this work,
smaller galaxies ($M \la$ 10$^9$-10$^{10} \Msun$) will form from 2-$\sigma$ 
peaks 
at typical redshifts $z_f \sim$ 4-5 and from (rarer) 3-$\sigma$ peaks 
at $z_f \sim$ 
6-8 and larger galaxies somewhat later (but not later than in CDM models as
$\sigma(M)$ is almost the same in the WDM1, WDM2 and CDM models for $M \ga
10^{11}$ \Msun).
For low $\Omega_M$ cosmologies the formation redshifts would be expected to
be even higher than this, so qualitatively the WDM scenario does not seem to 
be in trouble on this point. Moreover, Schaeffer \& Silk (\cite{SS88}) found
that warm dark matter can provide a galaxy distribution that is close to
observed galaxy counts, provided $\Gamma \equiv \Omega_M h >$ 0.1, which is
likely to be the case. In particular they found that in the warm dark matter
structure formation scenario small-scale objects ($M \la 10^{10}$ \Msun) are
not overproduced, contrary to what may be the case for CDM, cf. 
section~\ref{s:intro}.

A general problem for $\Omega_M$=1 models is major (mass ratios $\sim$ 1:4 or
more), late merging events, which
potentially can be fatal for the fragile, stellar disks. Such late merging
events also occur in our WDM simulations of galaxies S2-S4 at redshifts 
$z$=0.2-0.4. This provides yet another reason for going to low $\Omega_M$ 
models,
since such merging would be expected to take
place faster and earlier, leaving more time afterwards for the disks to grow 
smoothly and steadily from cooling flows and to form stars out of the cool gas.

Finally, from a particle physics point of view, there are plenty of candidates 
for warmons. Their properties and initial momentum distributions may be 
significantly different. They could be self-interacting or sterile, long-lived
or absolutely stable. 
It is even possible that the {\it same} particles form both cold and warm dark
matter.
If so, the exciting possibility exists, that 
compared to CDM one could have smaller, but non-zero power on mass
scales less than the WDM free-streaming scale ($\sim 10^{10}$-$10^{11}
M_{\odot}$). \\
To mention a few WDM particle candidates in summary, they could be neutrinos 
with weaker
or stronger couplings than normal, majorons (light pseudogoldstone bosons)
or shadow or mirror world neutrinos.  

In closing, given the success of WDM scenario in solving
the angular momentum problem and possibly other problems related to galaxy 
formation
and structure, N-body and more hydro/gravity simulations 
should be undertaken in various cosmologies to bring the WDM scenario on a
more rigorous footing.\\[1cm]

We have benefited from the comments of Per Rex Christensen, Rupert Croft,
Julien Devriendt,
Chris Flynn, Fabio Governato, Martin G{\"o}tz, Ben Moore, Bernard Pagel,
Massimo Persic, Elena Pierpaoli, Joel Primack,
{\"O}rn{\'o}lfur R{\"o}gnvaldsson, Jens Schmalzing, Joe Silk, Henrik
Vedel and the referees.
This work was supported by Danmarks Grundforskningsfond through its support
for the establishment of the Theoretical Astrophysics Center. AD thanks the
Issac Newton Institue for Mathematical Science for its hospitality during the
completion of this work.

\newpage

\figcaption[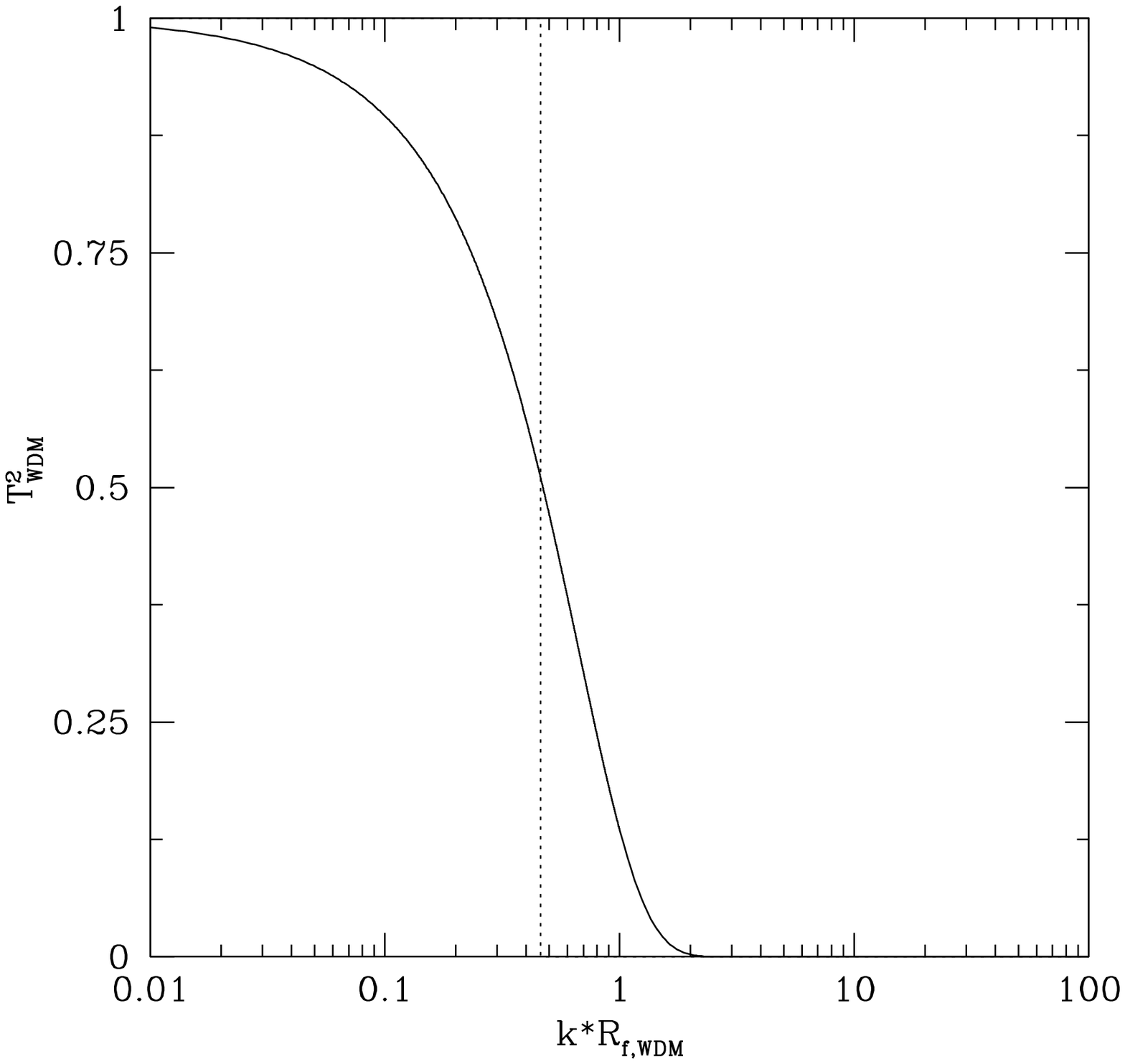]{The square of the CDM to WDM transfer function ({\it solid
line}) and the step-function approximation used in this work ({\it dotted 
line}).
\label{f:1}}

\figcaption[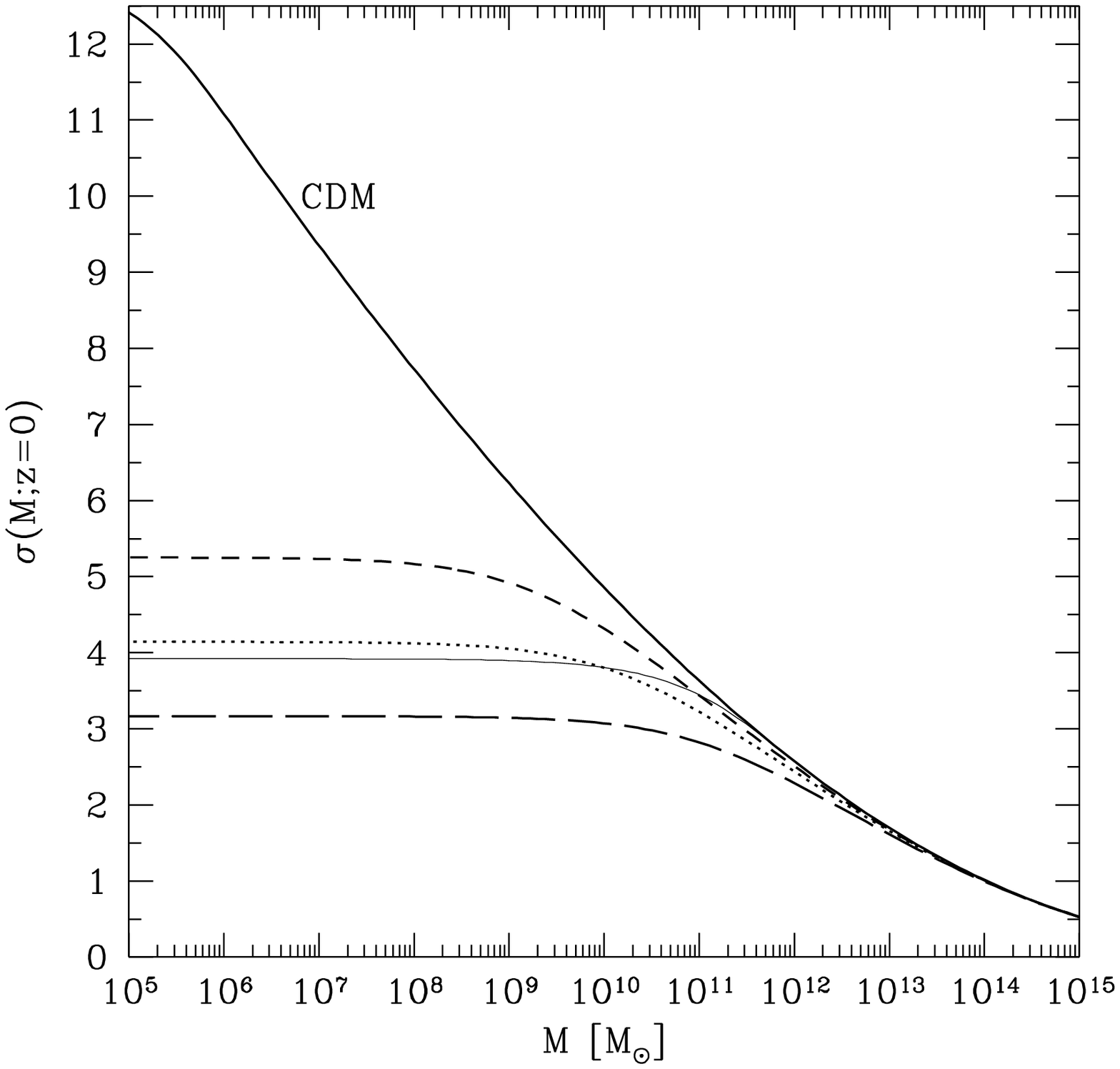]{The mass fluctuation dispersion $\sigma(M)$ at $z$=0
for CDM ({\it heavy solid line}) and WDM with $M_{f,WDM}=1.9
\times 10^{10} h^{-1} \Msun$ ({\it short-dashed line}), $1.5
\times 10^{11} h^{-1} \Msun$ ({\it dotted line}) and $1.2
\times 10^{12} h^{-1} \Msun$ ({\it long-dashed line}). Also shown is 
$\tilde{\sigma}(M;z$=0) for our step-function approximation to $T_{WDM}$
for $M_{f,WDM}=1.5 \times 10^{11} h^{-1} \Msun$ ({\it thin solid line}).
\label{f:2}}

\figcaption[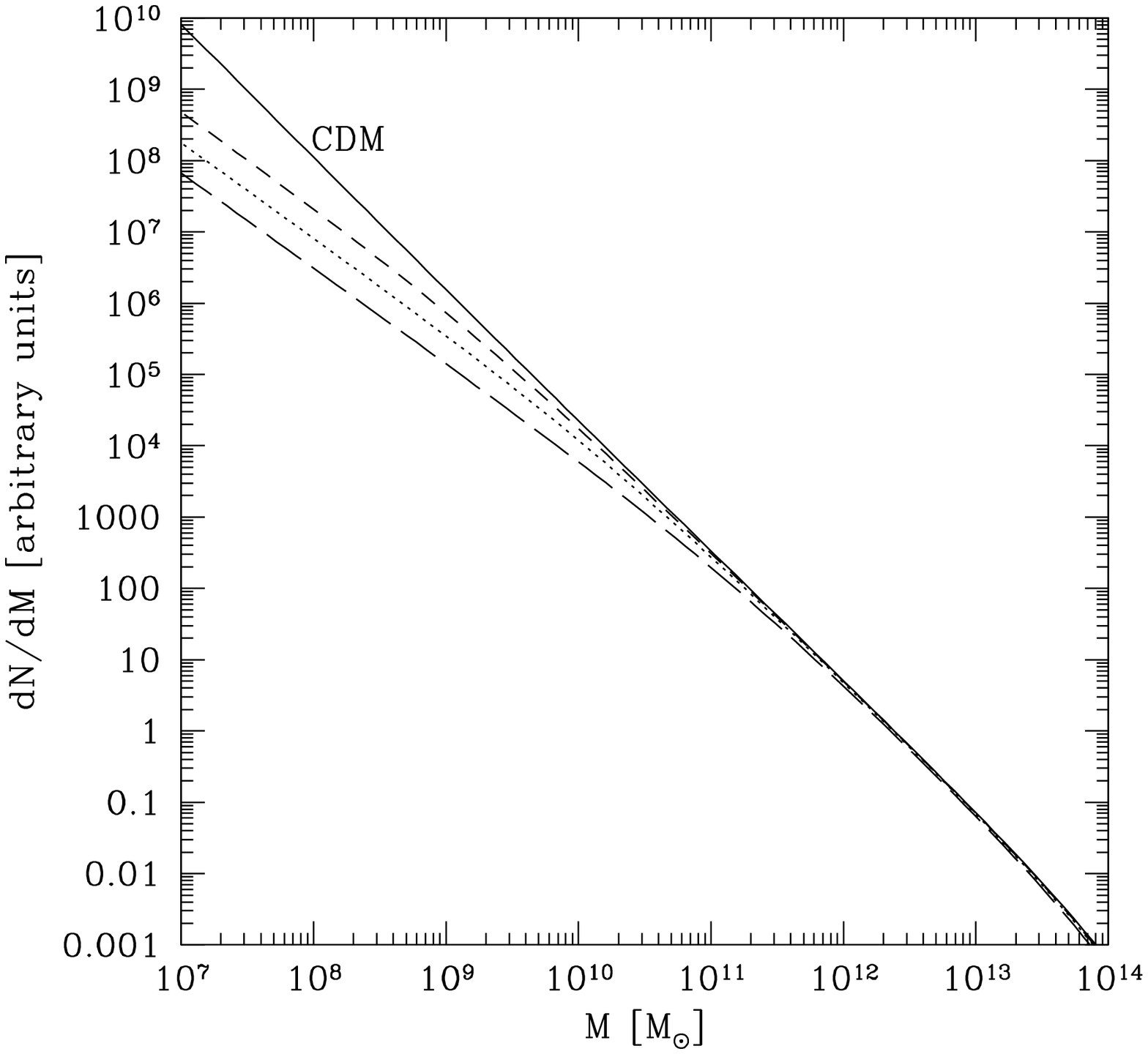]{Mass spectra $dN/dM$ at $z$=0 for CDM and the 
WDM from Fig.~\ref{f:2} (symbols as in Fig.~\ref{f:2}). 
\label{f:3}}

\figcaption[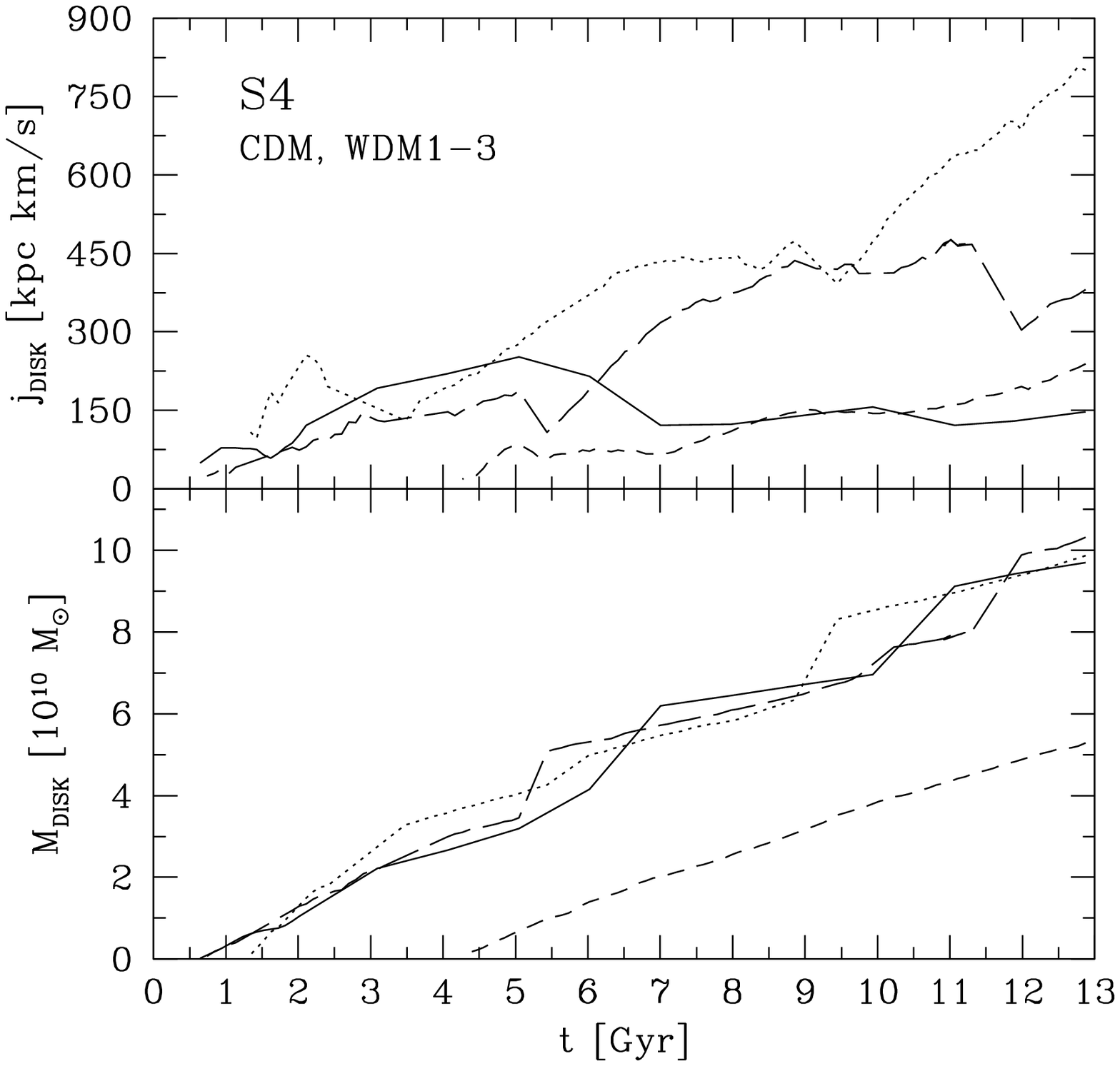]{Specific angular momenta $j_{disk}(t)$ and cooled-out
mass $M_{disk}(t)$ of the disk galaxies forming in the CDM and WDM1, WDM2, and 
WDM3, MR simulations (with $\Omega_b=0.05$ and UVX radiation field) of galaxy 
S4 - symbols as in Fig.~\ref{f:2}. 
\label{f:4}}

\figcaption[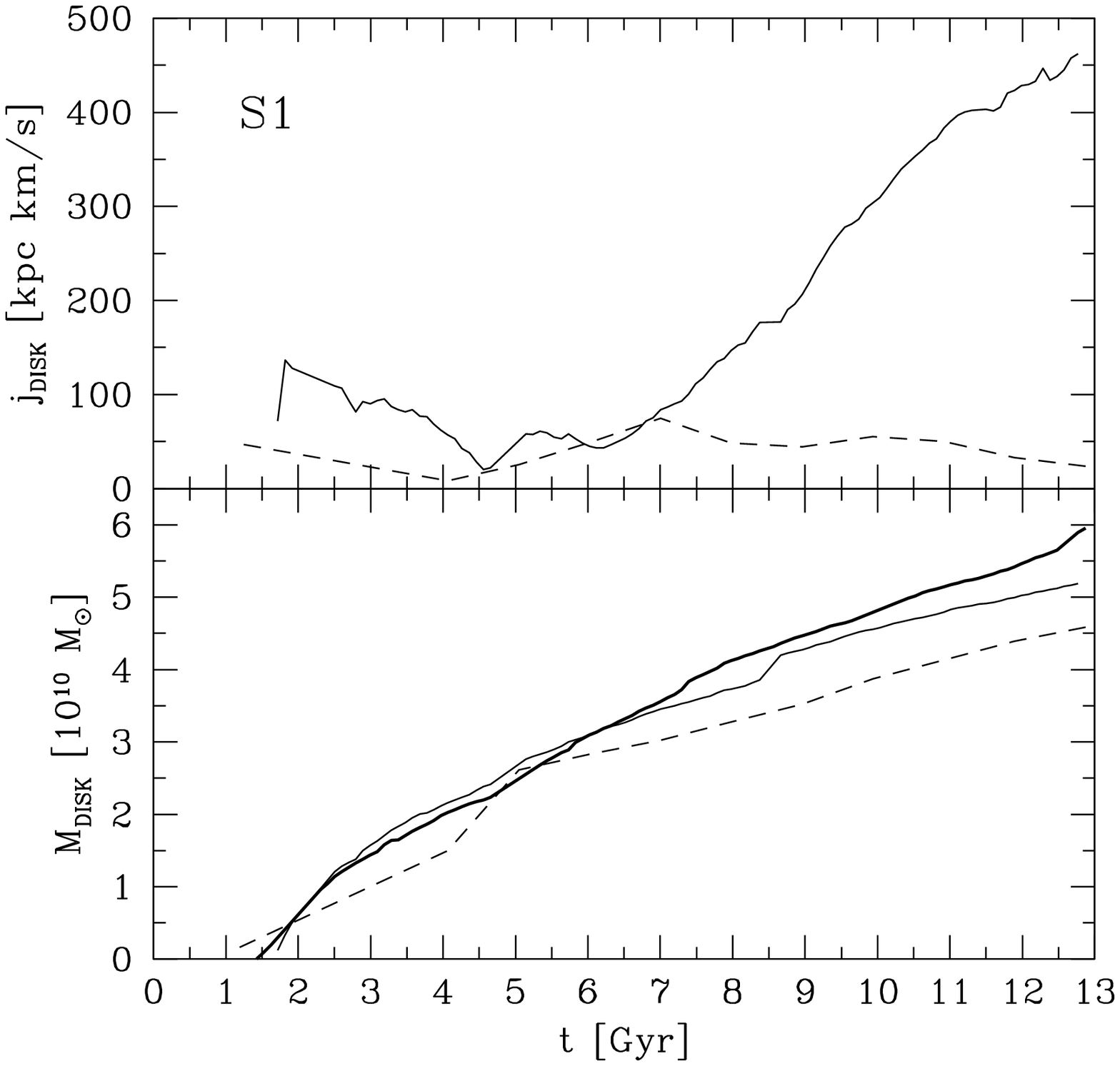]{Specific angular momenta $j_{disk}(t)$ and cooled-out
mass $M_{disk}(t)$ of the disk galaxies forming in the CDM ({\it dashed lines})
and 
WDM2, MR ({\it solid lines}) and HR ({\it heavy solid lines}) simulations 
(with $\Omega_b=0.05$ and UVX radiation field) of galaxy S1 ($j_{disk}(t)$ is 
not shown for the HR simulation of this galaxy - see text). 
\label{f:5}}

\figcaption[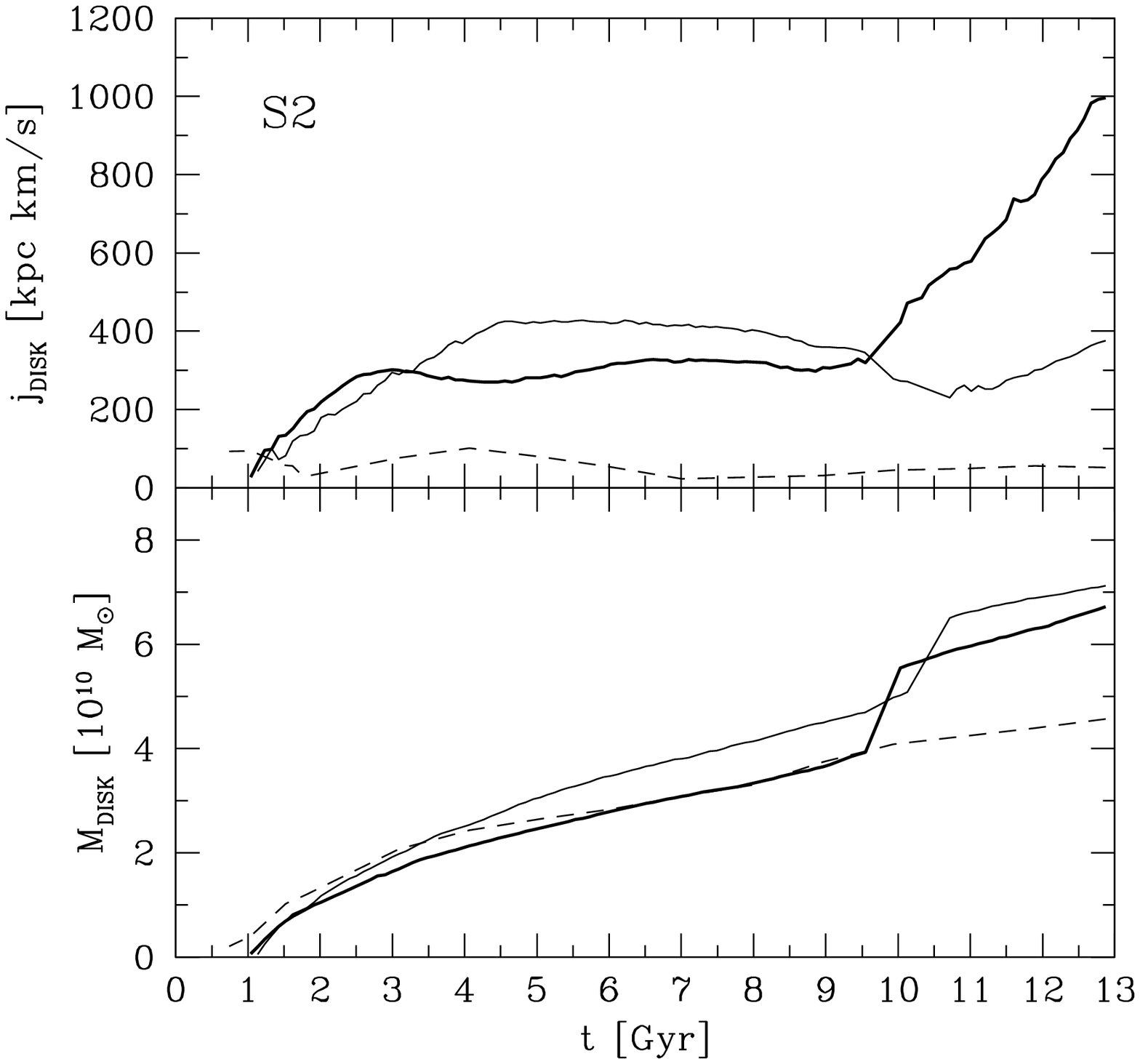]{Same as in Fig.~\ref{f:5}, but for galaxy S2. 
\label{f:6}}

\figcaption[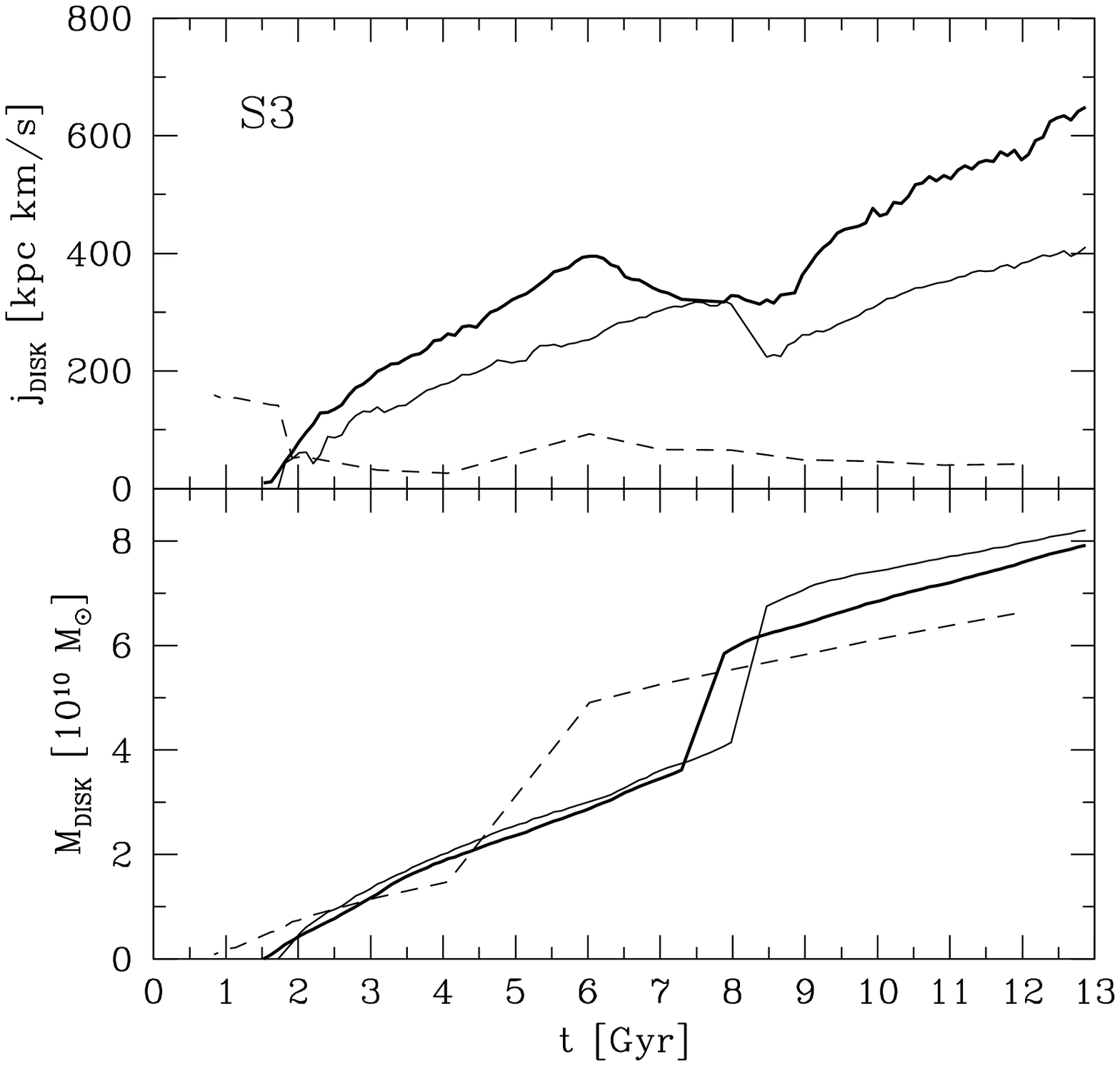]{Same as in Fig.~\ref{f:5}, but for galaxy S3. 
\label{f:7}}

\figcaption[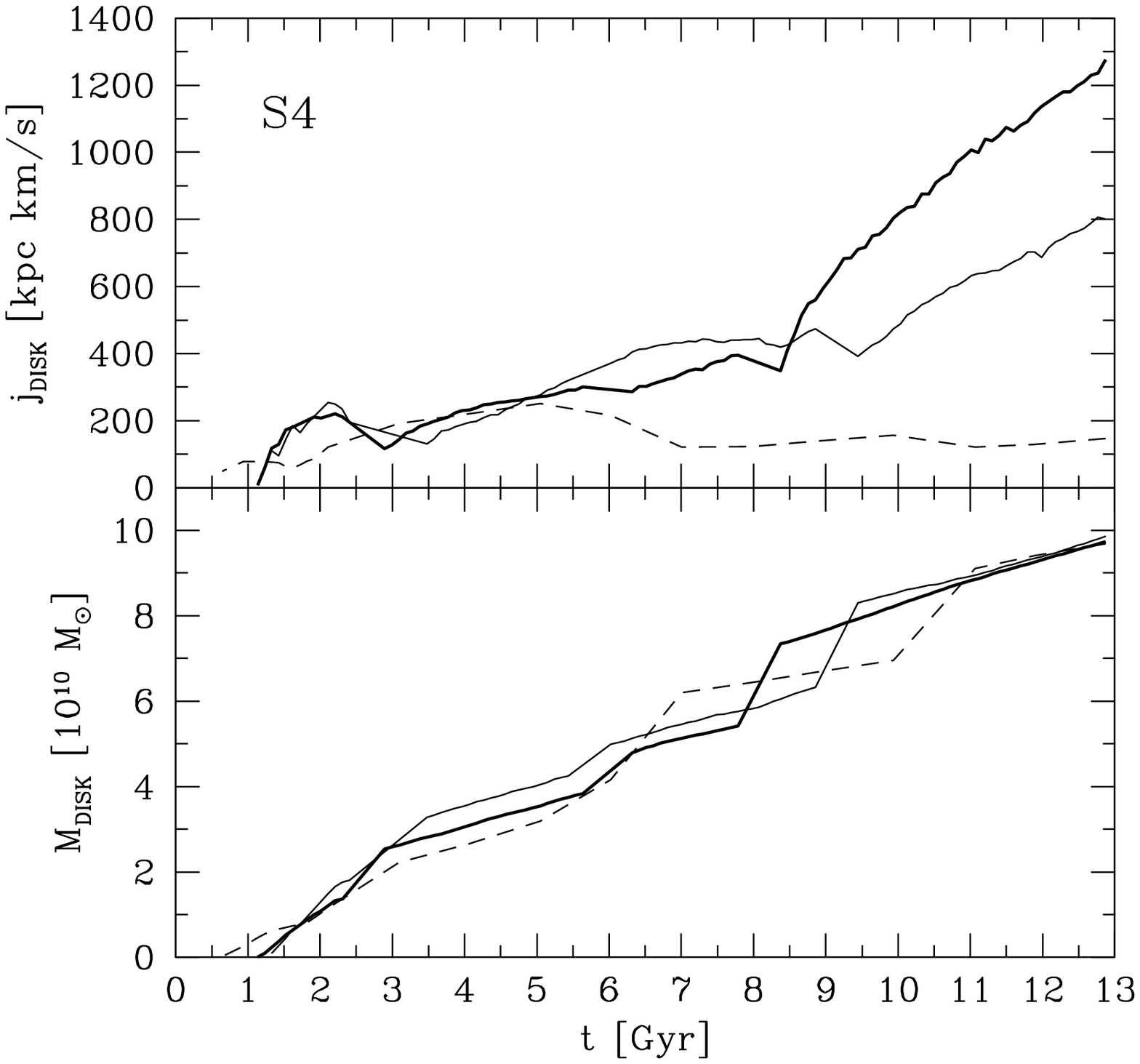]{Same as in Fig.~\ref{f:5}, but for galaxy S4. 
\label{f:8}}

\figcaption[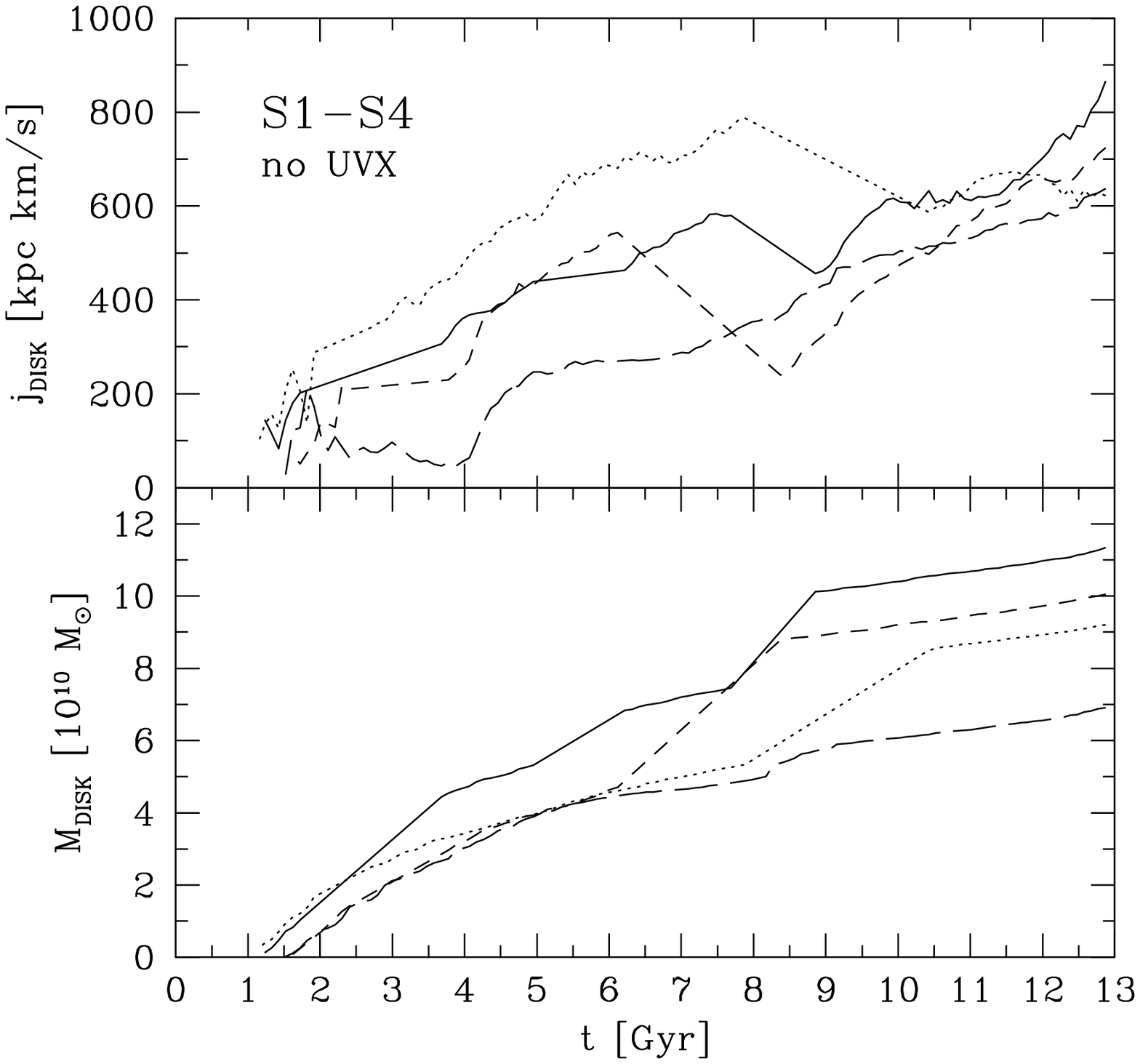]{Specific angular momenta $j_{disk}(t)$ and cooled-out
mass $M_{disk}(t)$ of the disk galaxies forming in the WDM2, MR simulations 
(with $\Omega_b=0.05$ and no UVX radiation field) of
galaxies S1 ({\it long-dashed line}), S2 ({\it dotted line}), S3 ({\it 
short-dashed line}) and S4 ({\it solid line}).
\label{f:9}}

\figcaption[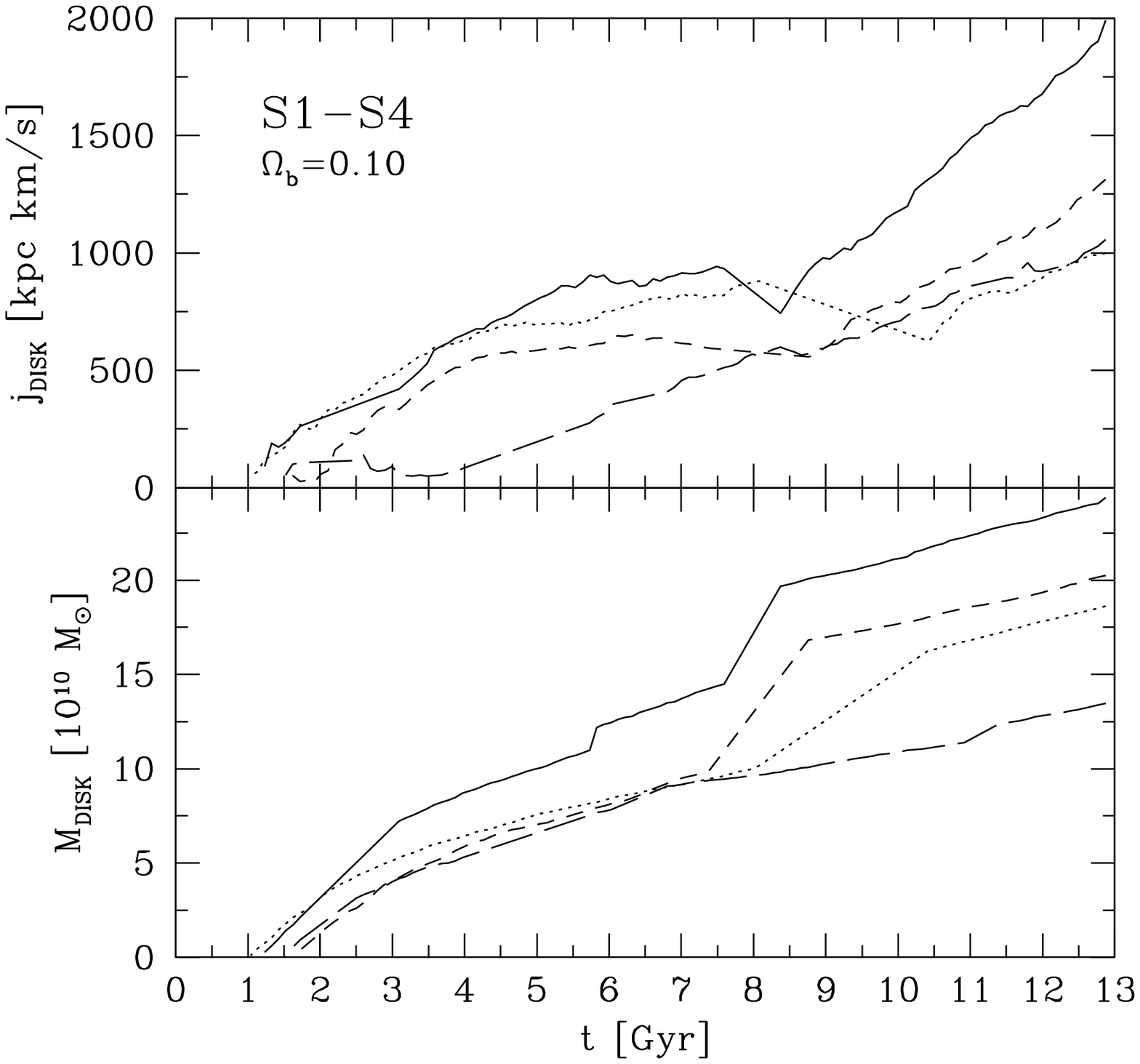]{Same as in Fig.~\ref{f:9}, but with $\Omega_b=0.10$ and 
UVX radiation field.
\label{f:10}}

\figcaption[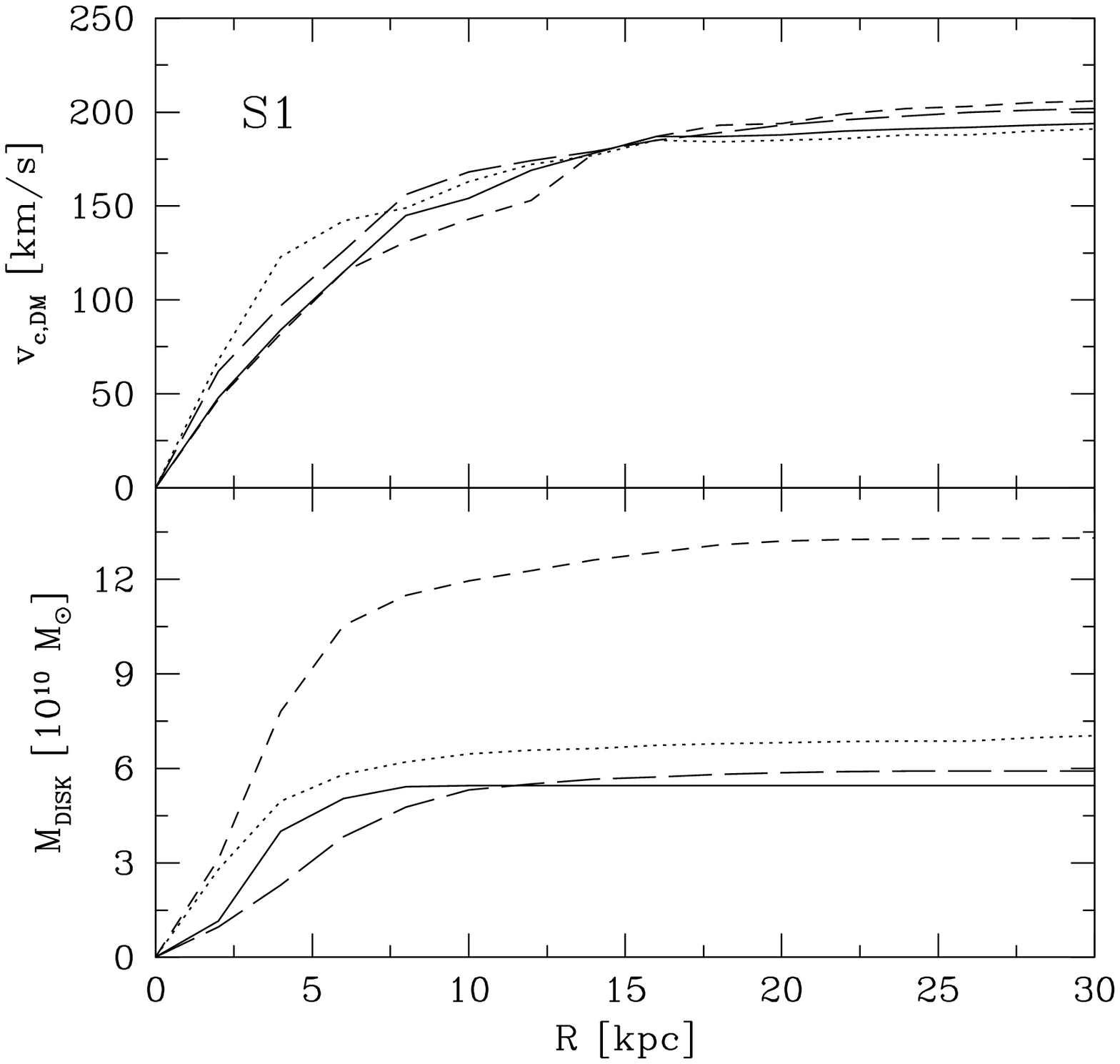]{Circular velocity $v_{c,DM}(R)$ of the final dark matter
halos and cooled-out disk mass $M_{disk}(R)$ of the final disk galaxies formed
in the WDM2 simulations of galaxy S1: The MR run with $\Omega_b=0.05$ and UVX 
radiation field ({\it solid lines}), the similar HR run ({\it long-dashed 
lines}), the MR run with $\Omega_b=0.05$ and no UVX radiation field ({\it
dotted line}) and the MR run with $\Omega_b=0.10$ and UVX radiation field 
({\it short-dashed lines}).
\label{f:11}}

\figcaption[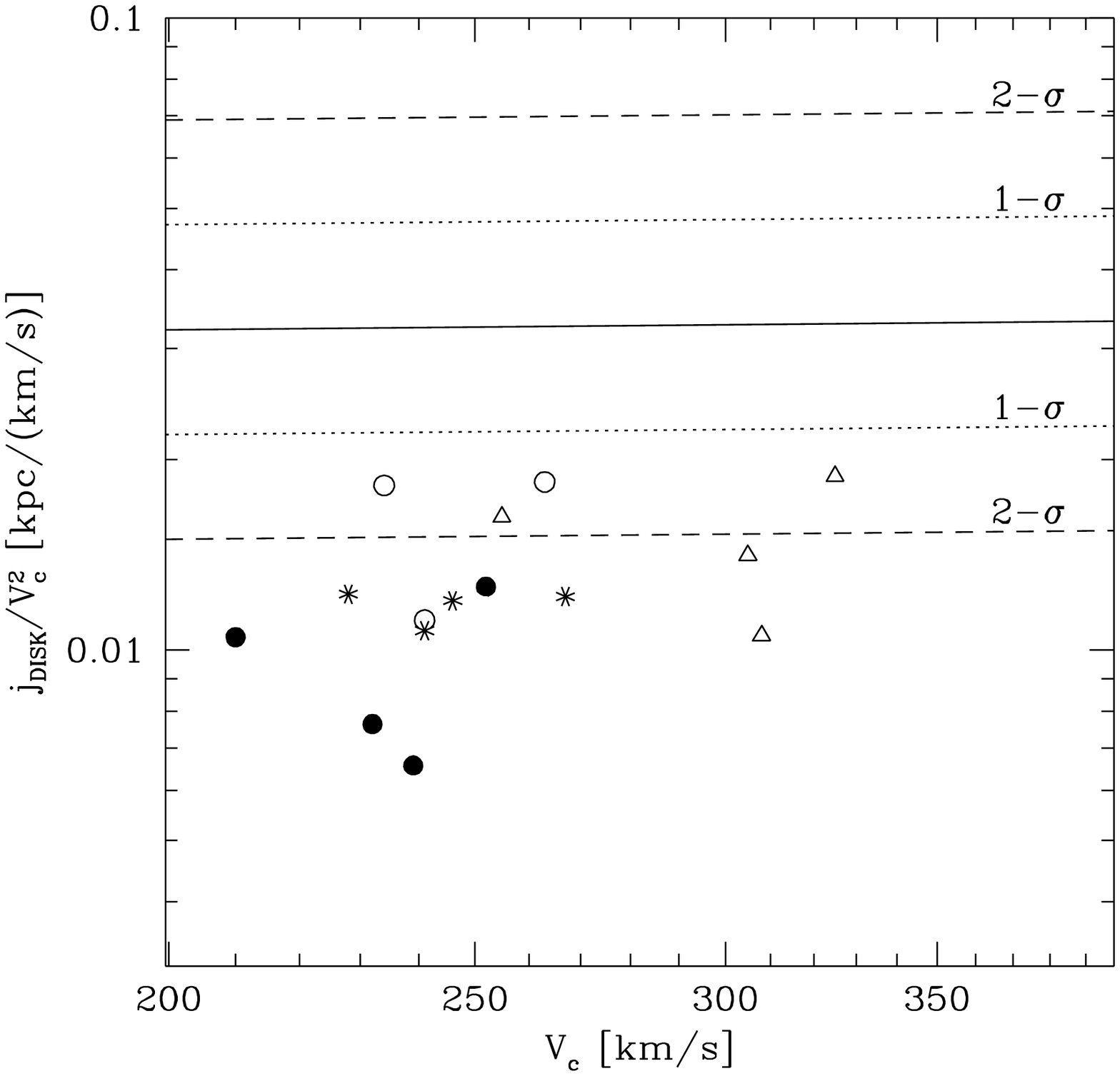]{Normalized specific angular momenta $\jt_\disk \equiv
j_\disk/V_c^2$ of the final disk galaxies formed in the WDM2 simulations:
The MR runs with $\Omega_b=0.05$ and UVX 
radiation field ({\it filled circles}), the similar HR run ({\it open circles};
S1 not shown - see text), the MR 
runs with $\Omega_b=0.05$ and no UVX radiation field ({\it asterisks}) and the 
MR runs with $\Omega_b=0.10$ and UVX radiation field ({\it open triangles}).
The solid line shows the median value from the
observational data, the dotted and dashed lines bracket the 1-$\sigma$
and 2-$\sigma$ intervals around this mean.
\label{f:12}}

\figcaption[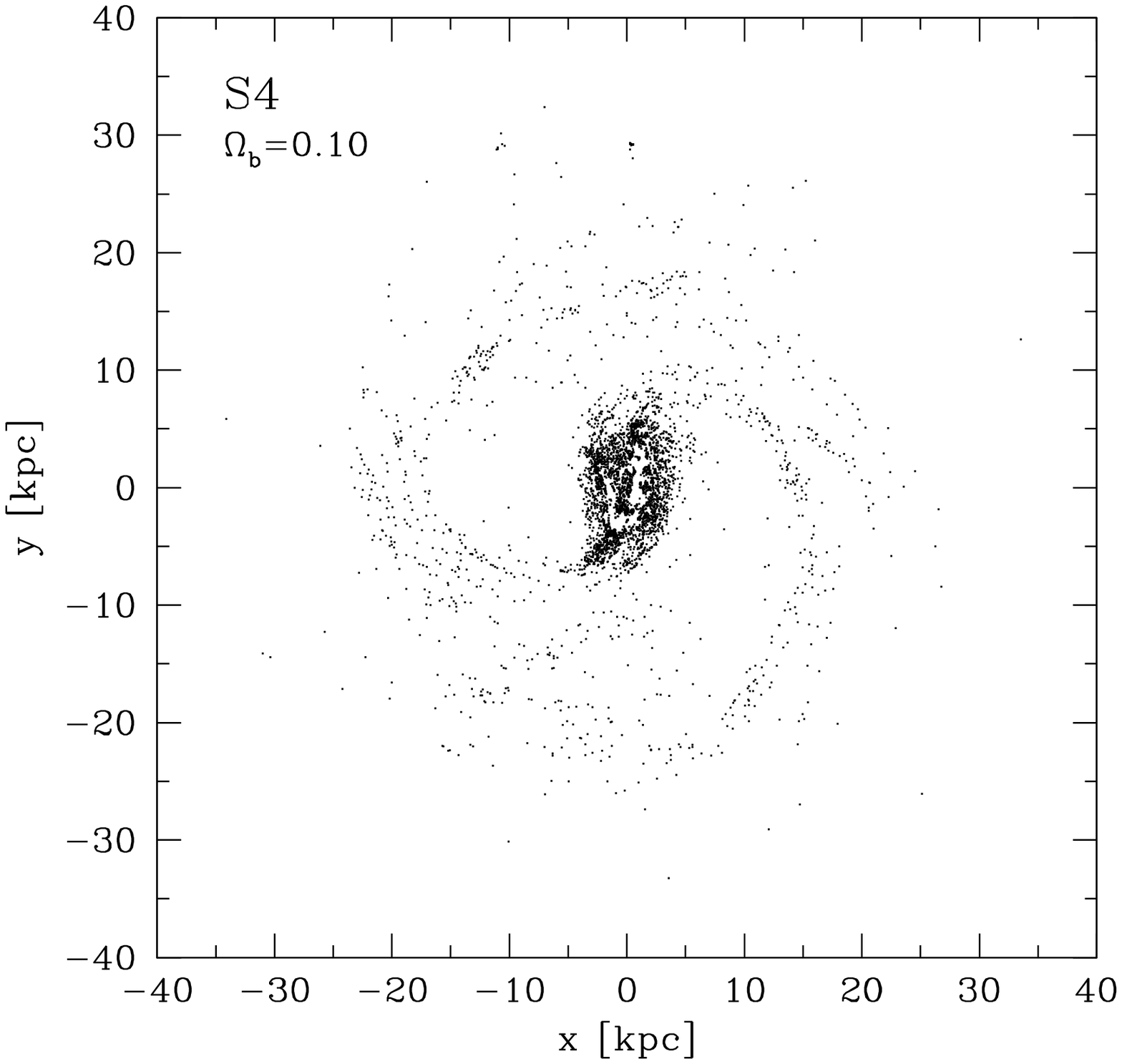]{Face-on view of the final galaxy formed in run \# 18 -
this disk galaxy has the largest specific angular momentum of all galaxies
formed in our WDM simulations, $j_{disk} \simeq$ 2000 kpc~km/s. Shown in
the Figure are all $\sim 4200$ SPH particles in the central disk with log($T) 
< 4.5$.
\label{f:13}}

\figcaption[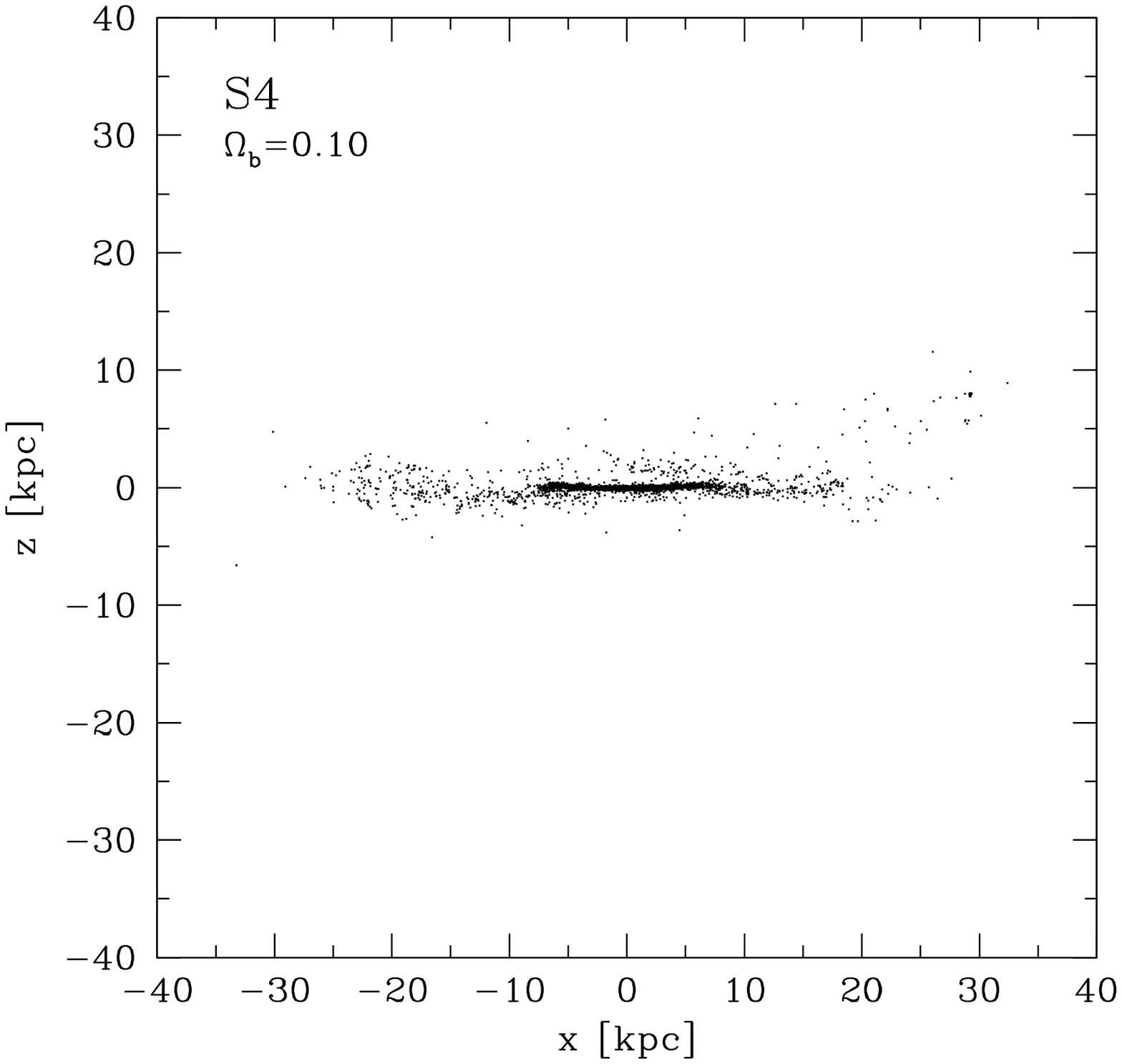]{An edge-on view of the galaxy in Fig.~\ref{f:13}. 
\label{f:14}}

\figcaption[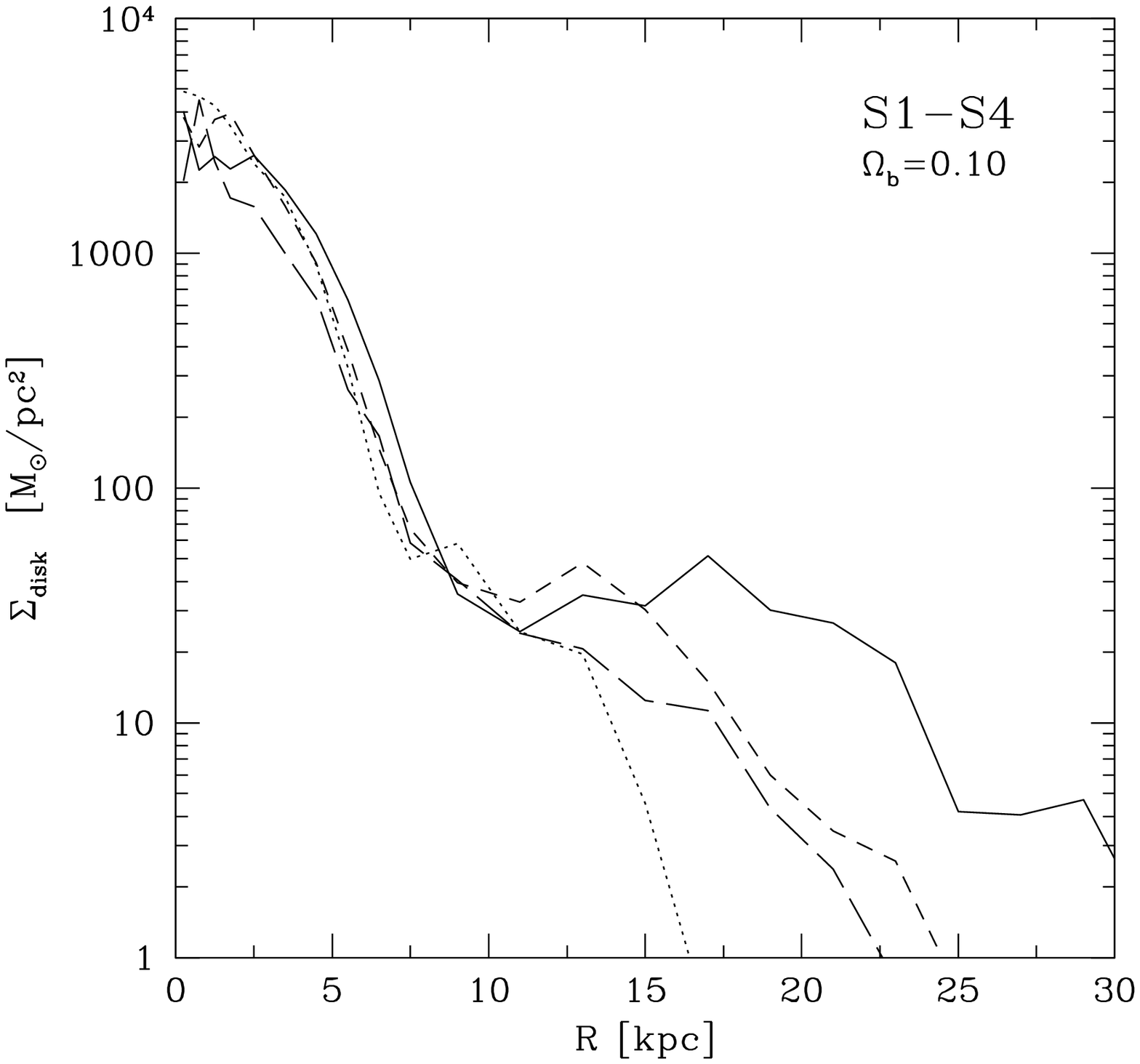]{Azimuthally averaged disk surface density profiles 
of the final disk galaxies formed in the WDM2, MR simulations 
(with $\Omega_b=0.10$ and UVX radiation field) of
galaxies S1 ({\it long-dashed line}), S2 ({\it dotted line}), S3 ({\it 
short-dashed line}) and S4 ({\it solid line}).
\label{f:15}}

\figcaption[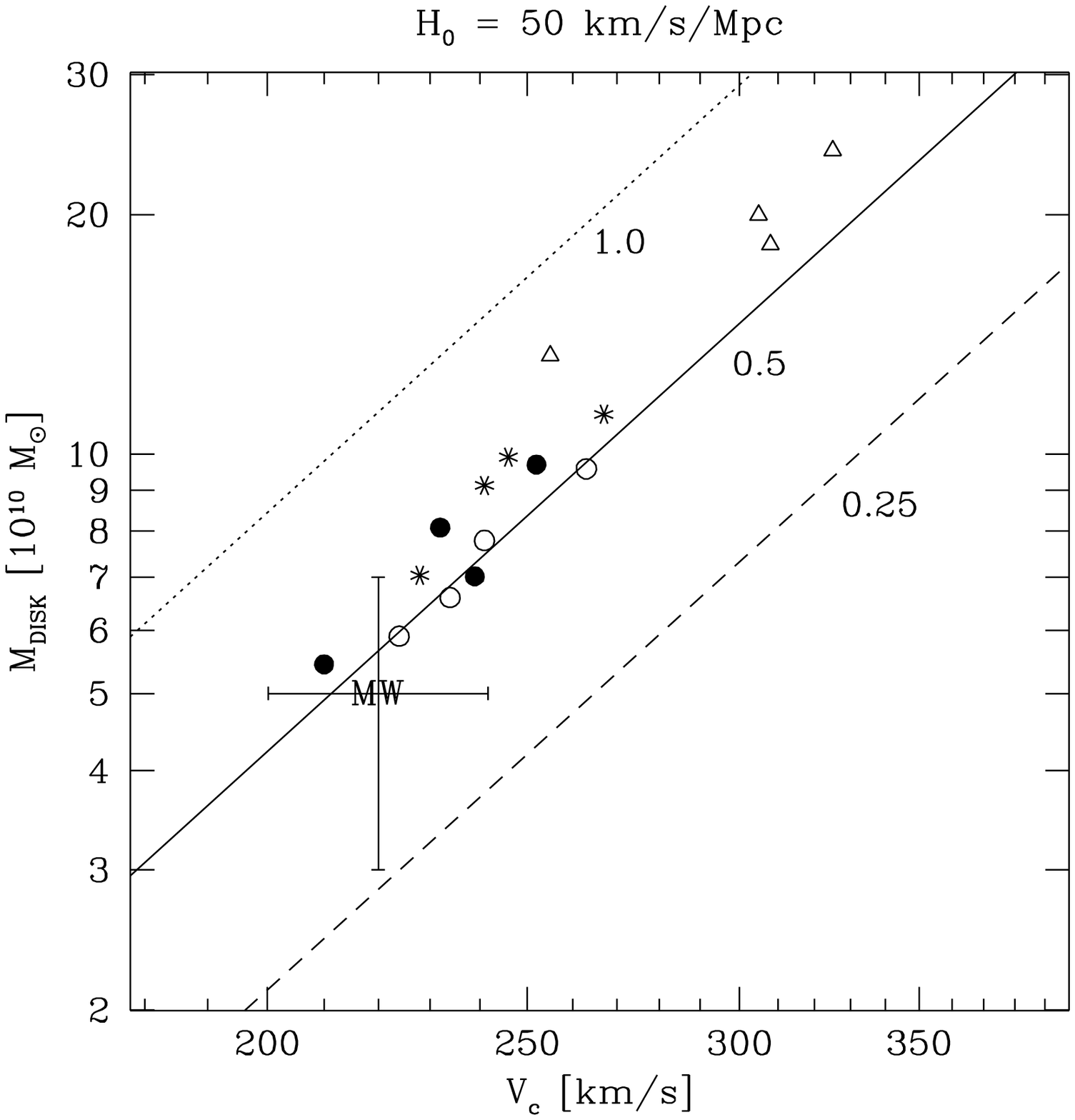]{The mass vs. circular velocity ``Tully-Fisher'' relation
for the final disks of our 16 WDM2 simulations - symbols used are the same
as in Fig.~\ref{f:12}. Also shown is the observed $I$-band TF relation of
Giovanelli \etal converted to mass assuming ($M/L_I$) = 0.25 ({\it dashed
line}), 0.5 ({\it solid line}) and 1.0 ({\it dotted line}). Finally, the
symbol ``MW'' with errorbars shows the likely range of the total, baryonic mass
and characteristic circular velocity of the Milky Way.
\label{f:16}}

\figcaption[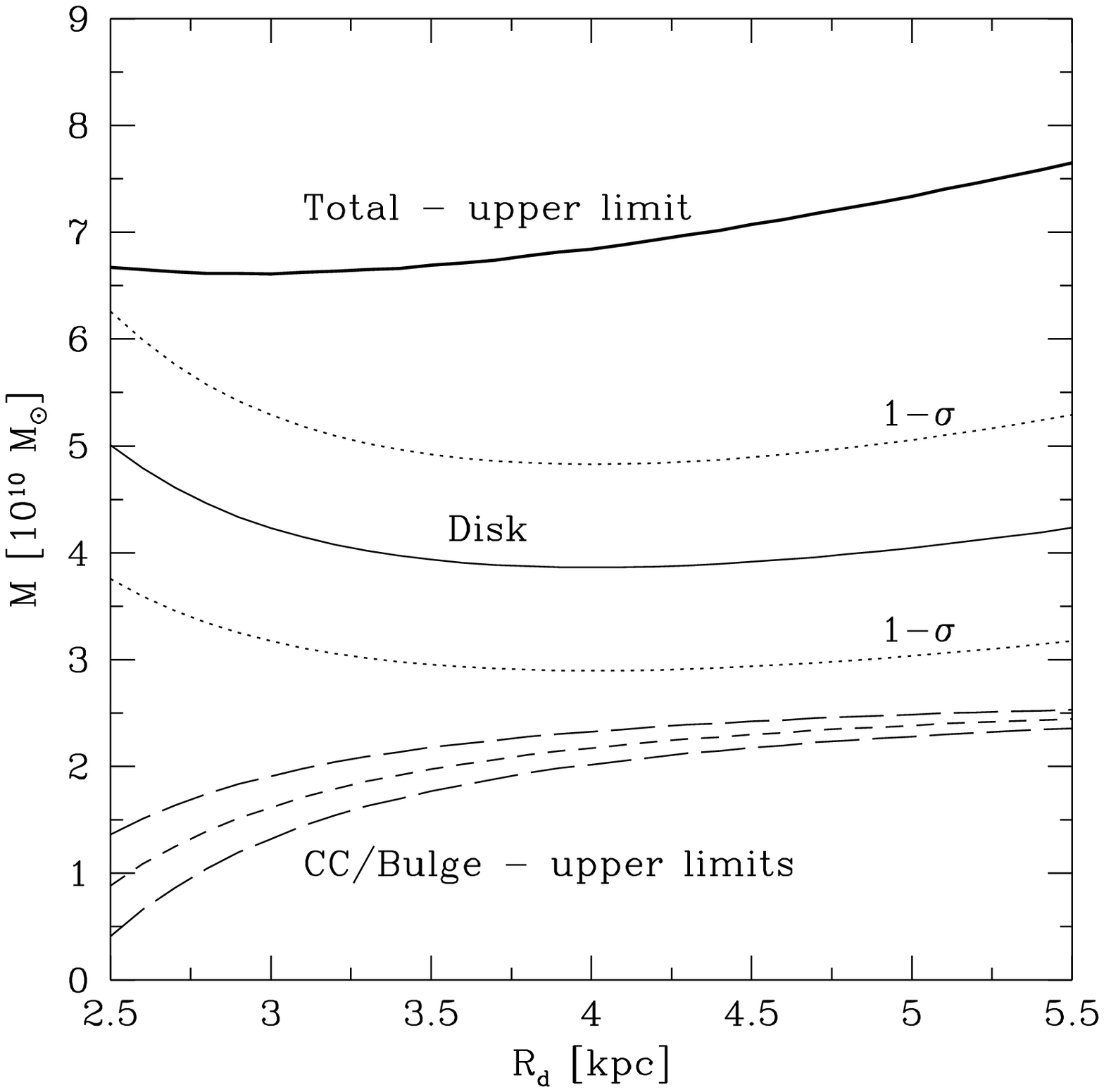]{Baryonic mass of the Milky Way as a function of the
disk scale-length $R_d$. The estimated mass of the disk is shown by the solid
line and the 1-$\sigma$ contours by the dotted lines. The upper limit to the 
mass of the central component/bulge is shown by the short-dashed line and
lower and upper 1-$\sigma$ contours by the long-dashed lines - see text for
details. The 1-$\sigma$ upper limit to the total, baryonic mass of the Milky 
Way is shown by the heavy, solid line.
\label{f:17}}

\clearpage

\newpage        
\begin{table}
\caption{Parameters of the simulations \label{t:np}}
\begin{tabular}{lcccccrr}
\hline\hline
Run & Galaxy & DM type & Resolution & UVX field & $\Omega_b$ & 
$N_{\rm SPH}$ & $N_{\rm DM}$\\
\hline
1 & S4 & WDM1 & MR & yes & 0.05 & 14736 & 7368\\
2 & S4 & WDM2 & MR & yes & 0.05 & 14700 & 7350\\
3 & S4 & WDM3 & MR & yes & 0.05 & 14848 & 7424\\
4 & S1 & WDM2 & MR & yes & 0.05 & 13700 & 6850\\
5 & S2 & WDM2 & MR & yes & 0.05 & 14142 & 7071\\
6 & S3 & WDM2 & MR & yes & 0.05 & 14404 & 7202\\
7 & S1 & WDM2 & HR & yes & 0.05 & 54689 & 54689\\
8 & S2 & WDM2 & HR & yes & 0.05 & 56386 & 56386\\
9 & S3 & WDM2 & HR & yes & 0.05 & 57723 & 57723\\
10 & S4 & WDM2 & HR & yes & 0.05 & 58739 & 58739\\
11 & S1 & WDM2 & MR & ~no & 0.05 & 13700 & 6850\\
12 & S2 & WDM2 & MR & ~no & 0.05 & 14142 & 7071\\
13 & S3 & WDM2 & MR & ~no & 0.05 & 14404 & 7202\\
14 & S4 & WDM2 & MR & ~no & 0.05 & 14700 & 7350\\
15 & S1 & WDM2 & MR & yes & 0.10 & 13700 & 6850\\
16 & S2 & WDM2 & MR & yes & 0.10 & 14142 & 7071\\
17 & S3 & WDM2 & MR & yes & 0.10 & 14404 & 7202\\
18 & S4 & WDM2 & MR & yes & 0.10 & 14700 & 7350\\
\hline \hline
\end{tabular}
\end{table}

\clearpage

\begin{table}
\caption{Masses, sizes and velocities at $z=0$ \label{t:mrv}}
\small
\begin{tabular}{lcccccccccc}
\hline\hline
Run  & $M_{200}$ & $r_{200}$ & $V_{200}$ & $N_{gas}$ & $N_{DM}$ &
$M_{gas}$ & $M_{DM}$ & $N_{disk}$ & $M_{disk}$ & $M_{disk}$ \\
                         & [10$^{12}$ M$_{\odot}$]  & [kpc] & [km~s$^{-1}$] & & &
                              [10$^{10}$ M$_{\odot}$]  & [10$^{12}$ M$_{\odot}$] & &    
                              [10$^{10}$ M$_{\odot}$]  & [$\Omega_b M_{200}$]       \\
\hline
1 & 3.25 & 382 & 191 & ~5450 & ~2850 & 15.59 & 3.10 & ~3807 & 10.89 & 0.67\\
2 & 3.39 & 388 & 194 & ~5663 & ~2969 & 16.20 & 3.23 & ~3392 & ~9.70 & 0.57\\
3 & 2.42 & 346 & 173 & ~4101 & ~2098 & 11.82 & 2.30 & ~1814 & ~5.19 & 0.43\\
4 & 2.16 & 334 & 167 & ~3590 & ~1895 & 10.27 & 2.06 & ~1903 & ~5.44 & 0.50\\
5 & 2.37 & 344 & 172 & ~4125 & ~2074 & 11.80 & 2.25 & ~2452 & ~7.02 & 0.59\\
6 & 2.74 & 360 & 181 & ~4812 & ~2391 & 13.77 & 2.60 & ~2824 & ~8.08 & 0.59\\
7 & 2.18 & 334 & 167 & 15573 & 15206 & 11.14 & 2.07 & ~8246 & ~5.90 & 0.55\\
8 & 2.38 & 344 & 173 & 16955 & 16662 & 12.13 & 2.26 & ~9224 & ~6.60 & 0.55\\
9 & 2.68 & 358 & 179 & 19326 & 18686 & 13.82 & 2.54 & 10884 & ~7.79 & 0.58\\
10 & 3.29 & 384 & 192 & 22602 & 23045 & 16.17 & 3.13 & 13392 & ~9.70 & 0.58\\
11 & 2.15 & 332 & 167 & ~3566 & ~1822 & 10.20 & 2.05 & ~2459 & ~7.04 & 0.70\\
12 & 2.38 & 344 & 172 & ~4429 & ~2074 & 12.10 & 2.25 & ~3191 & ~9.13 & 0.77\\
13 & 2.78 & 362 & 182 & ~4818 & ~2426 & 13.78 & 2.64 & ~3464 & ~9.91 & 0.71\\
14 & 3.44 & 390 & 195 & ~5801 & ~3013 & 16.60 & 3.28 & ~3922 & 11.22 & 0.66\\ 
15 & 2.16 & 332 & 167 & ~3543 & ~1892 & 20.27 & 1.95 & ~2338 & 13.38 & 0.62\\
16 & 2.42 & 346 & 174 & ~4357 & ~2110 & 24.93 & 2.17 & ~3203 & 18.33 & 0.76\\
17 & 2.83 & 364 & 183 & ~4981 & ~2475 & 28.50 & 2.55 & ~3486 & 19.96 & 0.70\\
18 & 3.46 & 390 & 195 & ~5945 & ~3025 & 34.02 & 3.12 & ~4205 & 24.06 & 0.69\\ 
\hline \hline
\end{tabular}
\end{table}

\begin{table}
\caption{Circular velocities and specific angular momenta at $z=0$ \label{t:j}}
\begin{tabular}{lcccc}
\hline\hline
Run  & $V_c$ & $\lambda$ & $j_{disk}$ & $r_{inf}$\\
                         & [km~s$^{-1}$] & & [kpc km~s$^{-1}$] & [kpc]\\
\hline
1 & 286 & 0.056 & ~381 & 176\\
2 & 252 & 0.045 & ~800 & 151\\
3 & 173 & 0.011 & ~239 & 120\\
4 & 210 & 0.026 & ~462 & 122\\
5 & 239 & 0.027 & ~375 & 145\\
6 & 232 & 0.045 & ~411 & 149\\
7 & 224 & 0.017 & - & 132\\
8 & 234 & 0.025 & ~997 & 134\\
9 & 241 & 0.049 & ~648 & 144\\
10 & 263 & 0.050 & 1276 & 150\\
11 & 228 & 0.018 & ~637 & 166\\
12 & 241 & 0.018 & ~623 & 199\\
13 & 246 & 0.046 & ~724 & 201\\
14 & 267 & 0.052 & ~866 & 178\\
15 & 255 & 0.020 & 1056 & 148\\
16 & 308 & 0.022 & 1000 & 201\\
17 & 305 & 0.045 & 1312 & 194\\
18 & 325 & 0.054 & 1990 & 186\\
\hline \hline
\end{tabular}
\end{table}

\end{document}